\documentclass[prb,aps,twocolumn, superscriptaddress, longbibliography,floatfix]{revtex4-2}
\usepackage[
  colorlinks=true,
  urlcolor=blue,
  linkcolor=blue,
  citecolor=blue,
  bookmarks=false,
  pdftitle={},
  pdfauthor={}
]{hyperref}
\usepackage{import}
\usepackage{subfiles}
\usepackage{graphicx}
\usepackage{bm}
\usepackage{amsmath}
\usepackage{amssymb}
\usepackage{mathtools}
\usepackage{amsfonts}
\usepackage[all]{xy}
\usepackage{xspace}
\usepackage{accents}
\usepackage{stmaryrd}
\usepackage{tabularx}
\usepackage{extarrows}
\usepackage{float}
\usepackage{multirow}
\usepackage{makecell}
\usepackage[capitalise]{cleveref}
\usepackage[dvipsnames]{xcolor}
\usepackage[normalem]{ulem}
\usepackage{soul}
\usepackage{pdfpages}
\makeatletter
\AtBeginDocument{\let\LS@rot\@undefined}
\makeatother

\usepackage{changes}
\definechangesauthor[color=Red]{SR}

\begin{document}

\title{Magneto tunnel conductance across twisted Weyl semimetal junctions}
\author{Nirnoy Basak}
\affiliation{Harish-Chandra Research Institute, A CI of Homi Bhabha National
Institute, Chhatnag Road, Jhunsi, Prayagraj (Allahabad) 211019, India}
\author{Sumathi Rao}
\affiliation{International Centre for Theoretical Sciences, Tata Institute of Fundamental Research, Bengaluru 560089, India}
\author{Faruk Abdulla}
\affiliation{Physics Department, Technion - Israel Institute of Technology, Haifa 32000, Israel}
\affiliation{The Helen Diller Quantum Center, Technion, Haifa 32000, Israel}

\begin{abstract}
We investigate magnetotransport across an interface between two  Weyl semimetals
(finite in both directions) whose Weyl 
nodes project onto two different surfaces which are twisted with respect to each other before being 
coupled. This gives rise to a novel contribution to the conductance through the junction purely 
through Fermi arc states, even in the absence of a magnetic field perpendicular to the junction. 
When the perpendicular magnetic field is included, we find that for a mesoscopic or smaller samples,  
the transverse Fermi arc states have a significant contribution to the conductance for experimentally 
relevant fields, and need to be taken into account along with the conductance through the bulk chiral 
Landau levels.
\end{abstract}

\maketitle

\section{Introduction}

Weyl semimetals (WSMs) \cite{Murakami_2007, Wan_Savrasov_2011, Yang_Ran_2011, 
Burkov_Balents_2011, Xu_Fang_2011, Lv_Ding_2015a, Lv_Ding_2015b, Xu_Hasan_2015a, 
Xu_Hasan_2015b, Lu_Soljacic_2015} are a class of three-dimensional topological materials 
in which non degenerate valence and conduction bands touch each other  at isolated points 
(called Weyl nodes) in the bulk Brillouin zone (BZ). The fact that the Weyl nodes 
in these  materials  carry non zero Chern numbers and act as  a source/sink 
of Berry curvature in the momentum space leads to a variety of phenomena - $e.g.$, 
 the chiral anomaly in the presence of parallel electric and magnetic fields, the negative 
magneto resistance, the planar Hall effect, etc. \cite{Nielsen_Ninomiya_1983,  Aji_2012, 
Zyuzin_Burkov_2012, Son_Spivak_2013, Gorbar_Miransky_2014, Burkov_2015,  
Lu_Shun_2017, Nandy_Tewari_2017,Das_Singh_2020,  Li_Shen_2018, Shama_Singh_2020,  
Li_Yao_2023,  Wei_Weng_2023}.

Due to the nontrivial topology of the Weyl nodes in the bulk, the WSMs host a special kind of 
surface states called Fermi arc surface states \cite{Wan_Savrasov_2011, Xu_Hasan_2015a, 
Xu_Hasan_2015b} joining the projections of 
Weyl nodes of opposite chirality on  the surface BZ. These Fermi arcs on the surface,  
in the 
presence  of bulk chiral states,   are known to exhibit thickness dependent quantum 
oscillations which provide  an observable signature of the chiral and topological 
character of Weyl nodes in  WSMs \cite{Potter_Vishwanath_2014, Zhang_Vishwanath_2016, 
Moll_Analytis_2016}. In a finite  system, the  Fermi arc surface 
states can also lead to nontrivial electronic transport \cite{Koshino_2016, Kaladzhyan_2019, 
Breitkreiz_2019, Sukhachov_2020}.  
 
The interface connecting two WSM  slabs, whether they are distinct or identical, possesses 
Fermi arc states that are confined to the junction \cite{Dwivedi_2018, Ishida_2018, Murthy_2020, 
faruk2021farecon, Mathur2023}. When the two WSM slabs are identical \cite{Murthy_2020, faruk2021farecon}, a 
slight  angular displacement is required at the junction to  have Fermi arc states at all. 

\begin{figure}[tbph]
 \centering
 \includegraphics[width=0.8\linewidth]{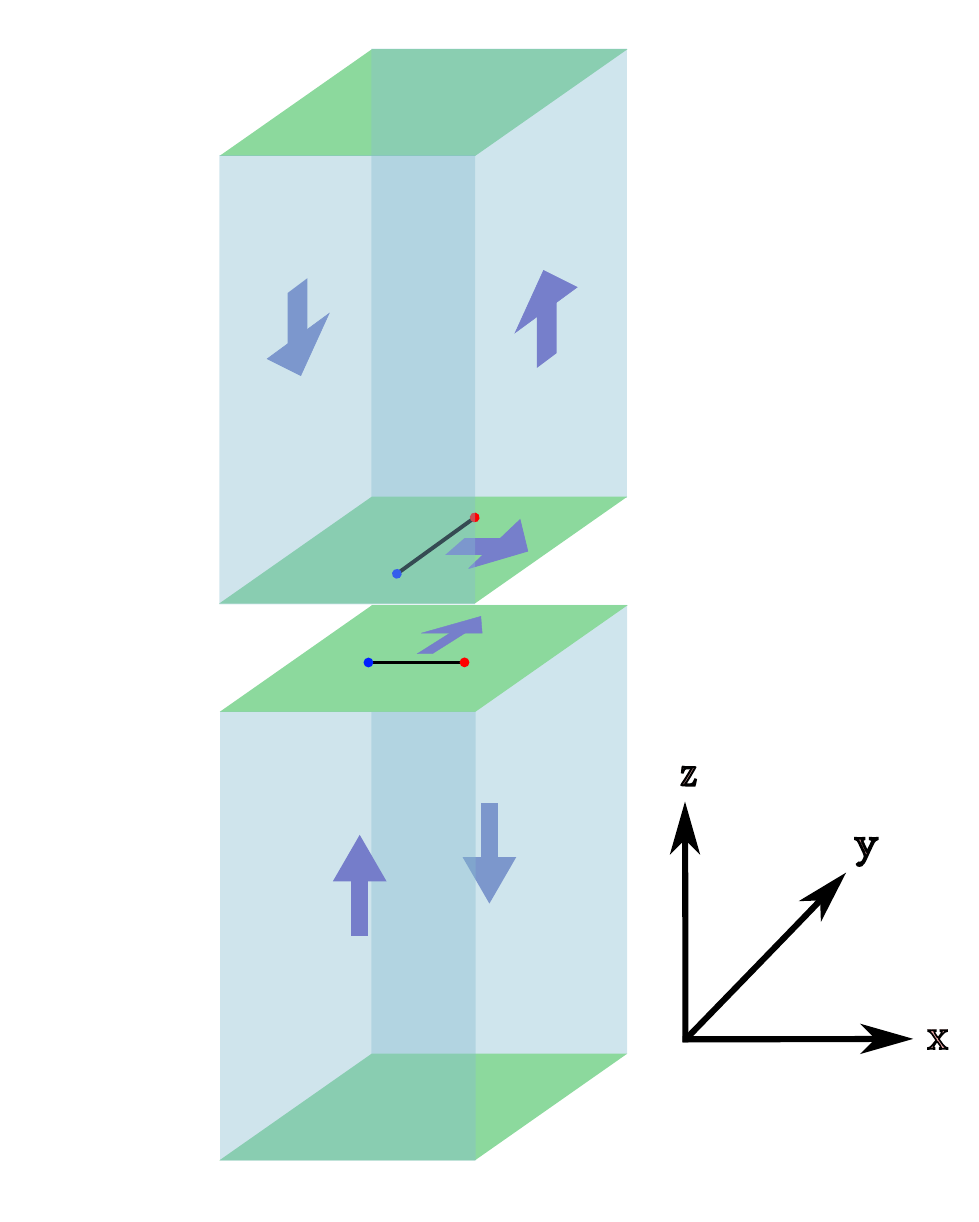}
 \caption{Schematic of the transverse surface Fermi arc states living in the $x$-$z$ plane (bottom slab) and in the $y$-$z$ plane (top slab). The blue arrows 
 signify the direction of the velocity of the transverse surface Fermi arc states on the 
 open surfaces. The dots represent the projection of Weyl nodes on the $k_xk_y$
 surface BZ, and the straight lines represent the Fermi arcs at the interface 
 at zero tunnel coupling. }
 \label{fig:sideways}
\end{figure}

These  Fermi arcs, confined to the junction, can bridge either the projections of Weyl nodes with 
the same chirality (homochiral connectivity) from the  two different WSM slabs or those with opposite 
chirality (heterochiral connectivity) from the same WSM slab. Since the Fermi arc states are localized 
at the junction (at $z=0$ as shown in Fig.\ref{fig:sideways}), it is  naively expected that
they cannot mediate current from one slab to the other ($i.e.$, along the $z$-direction) - their 
velocities are purely in the $xy$ plane. 
But in the presence of magnetic fields, electrons on the Fermi 
arcs can slide along the arc due to the Lorentz force  and  can transmit (or reflect)  current from one slab to another via the bulk states.
Recently, it was shown in Refs.\cite{chau2023magnetic, chaou2023quantum}  that in  the presence of a 
weak magnetic  field perpendicular to  the interface, the current   carried by the chiral Landau level is  either 
fully transmitted from  one slab to the other through the interface Fermi arcs if the  connectivity is
 homochiral,  or  it is fully reflected if the connectivity is  heterochiral.

However, when the WSM slabs are finite in the $x,y$ directions as well, Fermi arc states can also exist in the directions perpendicular or transverse to the junction (in the $xz$ or $xy$ planes, as shown in Fig.\ref{fig:sideways}). 
Here, the currents carried by the Fermi arc states on the transverse surfaces can also be transmitted through the interface Fermi arc states from one slab to the other \cite{buccheri2022transport}.  Earlier work 
\cite{chau2023magnetic, chaou2023quantum} ignored  the  contribution of  these states to  the tunnel conductance and focussed   only on  the 
contribution due to the bulk chiral Landau level states. 
The main focus of this work is to take into account 
the contribution due to these transverse Fermi arc states on the tunnel  conductance through the junction. We find that the current carried by these states are transmitted by the 
interface Fermi arcs from one slab to the other, without requiring any finite magnetic 
fields perpendicular to the interface.  In the absence of magnetic fields, we find that the 
transmitted current across the junction depends on the structure of the interface Fermi arcs localized 
at the junction.  The transmitted current  is maximised  when  the velocity of the electrons on the 
interface Fermi arc joining the  projections of Weyl nodes of the same chirality from  the two 
slabs are aligned (i.e. when the Fermi arc is a straight line). This  is when the hybridization of 
the Fermi arc states of the top slab and bottom slab gets maximised.

In the presence of magnetic fields perpendicular to the interface, previous work \cite{chau2023magnetic} 
found that the transmitted current mediated by the interface Fermi arc was directly proportional to the degeneracy of the chiral Landau levels and increased 
linearly with $B$ for small fields.  It was then shown to gradually saturate at larger fields due to backscattering between the Fermi arcs at the junction (referred to  as magnetic breakdown). 
In contrast, we find that when the transverse surface states are included, the total transmitted current is no longer linear with $B$ for small fields, because the number of modes in the surface channel is totally independent of the magnetic field. While for macroscopic samples, the surface contribution is small, for mesoscopic samples with length scales of O(10 microns), the surface contribution is higher than the contribution of the bulk chiral levels for magnetic fields
$ B < B_{crit}\sim 2 ~ {\rm Tesla}$. As the size of the sample increases, the critical field decreases.

For numerical simulations, we take two identical WSM slabs which are twisted by 
an angle $\pi/2$ about the  $z$-axis  and tunnel-coupled to form a junction at 
$z=0$. The rotation by an angle $\pi/2$ ensures that the projections of the Weyl nodes of 
one slab  are far away from the projections of the Weyl nodes of the other slab, 
so that the bulk Fermi surfaces of the two slabs do not overlap at small energies. 
Consequently,  there will be no direct current from one slab to the other 
\cite{tchoumakov201conservation, sousa2021gigantic, buccheri2022transport}. 
This enables us to unambiguously  study currents which are mediated by 
the localized Fermi arc states at the junction.

The plan of the paper is as follows: In Sec. II, we review interface  Fermi arcs at the 
junction  between the  two identical WSM slabs with a  relative twist angle of $\pi/2$. 
In  Sec. III,  we study electronic transport between the two slabs across the junction,  
in the absence of magnetic fields.  Then in Sec. IV, we study magnetotransport - 
transport mediated by the interface Fermi arcs, in the presence of a magnetic field which is perpendicular to the interface. 
Finally, we  discuss our findings  and summarize them in Sec. V. 

\section{Model and the interface Fermi arcs}

\begin{figure}[tbph]
	\centering
	\includegraphics[width=1.0\linewidth]{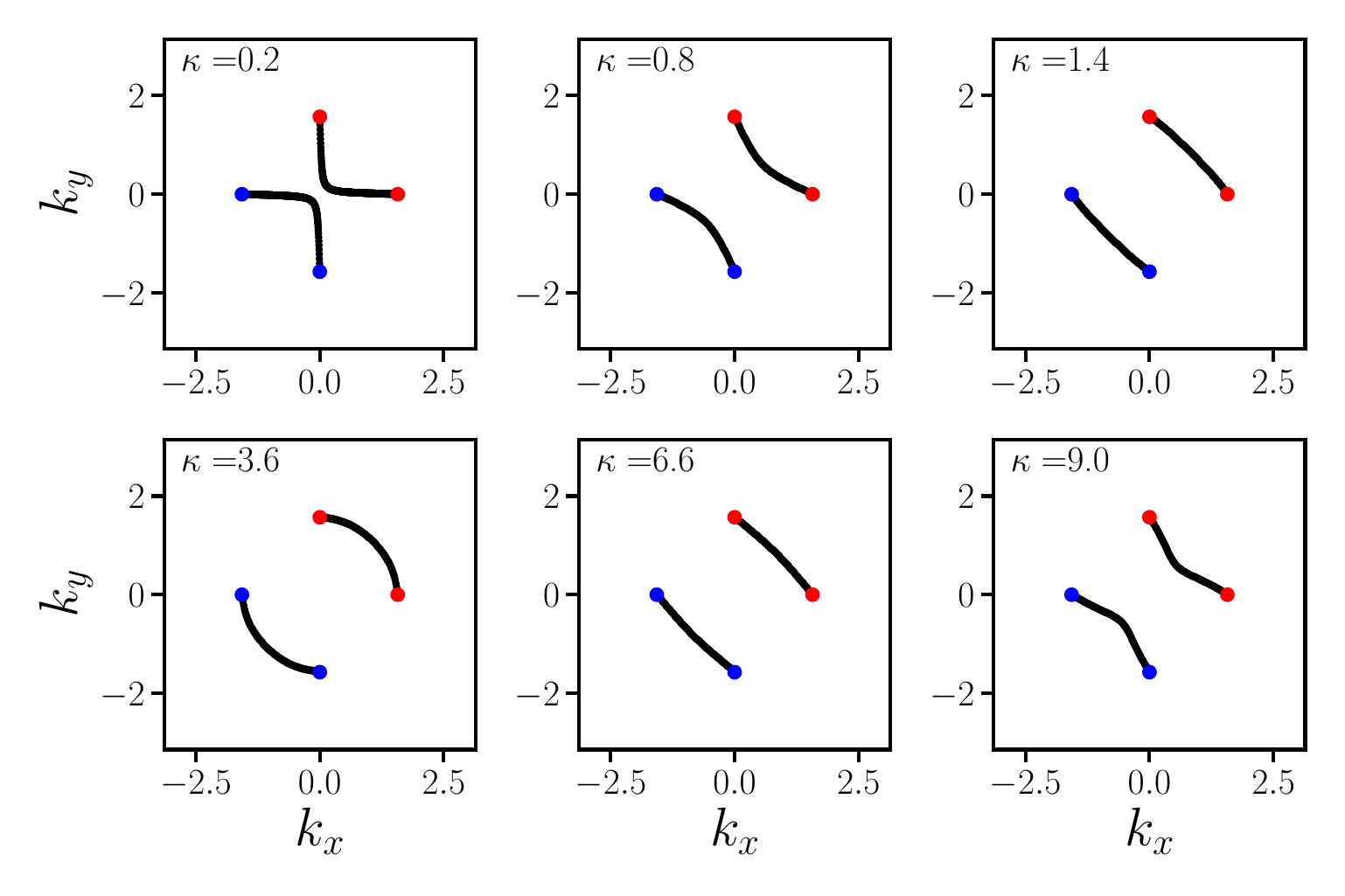}
 \caption{Plots of the interface Fermi arcs in the interface Brillouin zone. The Fermi arcs are plotted for 6 different values of the tunnel coupling($\kappa$). For all the plots $u=0.5$. The red and blue  dots are,  respectively, the projections of the positive and negative chirality Weyl nodes for both the  slabs on the interface BZ. It is clear from the plots that  the interface Fermi arcs connect projections of the  Weyl nodes of the same chirality. }
 \label{fig:reconfa}
\end{figure}

In this section, we describe our model and  discuss the Fermi arc states 
which are localized at the junction between two Weyl semimetals.  We consider 
two slabs of the same WSM with  two Weyl points each.  
The WSMs are finite along 
the $z$-direction with a thickness given by  $L_z$. The top and bottom slabs of the WSM, which are 
described by the Hamiltonian $H_t$ and $H_b$,  lie in the  regions $0 \le z \le L_z$ 
and $-L_z \le z \le 0$ respectively (see Fig. \ref{fig:sideways}). The slabs host Fermi 
arc surface states on those open surfaces on which projections of Weyl nodes of 
opposite chirality do not overlap. The two slabs are tunnel coupled at 
$z=0$: The top layer of the bottom slab is coupled to the bottom layer of the 
top slab. The full system is described by the following Hamiltonian 
\begin{align}
H = H_t + H_b + \kappa H_c, 
\end{align}
where the tunnel coupling term $H_c$ describes a hopping between the top layer 
of the bottom  slab and bottom layer of the top slab, and $\kappa$ parametrizes 
the strength of the tunnel coupling. The Hamiltonian $H_{\alpha}$ ($\alpha=t, b$) 
of the slabs are 
\begin{equation}
\begin{aligned}
H_{\alpha} = \sum_{n} & \sum_{{\bf k}}  c^{\dagger}_{\alpha, n}({\bf k}) \left(f_{\alpha1}({\bf k}) \sigma_x + 
f_{\alpha 2}({\bf k}) \sigma_z \right) c_{\alpha, n}({\bf k}) \\ &  -  \left(c^{\dagger}_{\alpha, n+1}({\bf k}) 
T c_{\alpha, n}({\bf k}) + H.c \right), 
\end{aligned}
\end{equation}
where ${\bf k} = (k_x, k_y)$,  $f_{b1}({\bf k}) = 2(2+\cos{k_0} - \cos{k_x} -\cos{k_y})$, 
$f_{b2}({\bf k})= 2\sin{k_y}$,  $f_{t1}({\bf k}) = 2(2+\cos{k_0} - \cos{k_y} - \cos{k_x} )$,
$f_{t2}({\bf k})=-2\sin{k_x}$, and the hopping matrix along the $z$-direction is 
$T = (\sigma_x + i\sigma_z)$. Here  $n$ represents the layer index along the 
$z$-direction.  The Hamiltonian $H_b$ of the bottom slab describes a 
time-reversal broken WSM with two Weyl nodes at ${\bf k}_w = \pm(k_0, 0, 0)$. 
We choose  $k_0=\pi/2$ so that the projections of the Weyl  nodes on the interface BZ 
are at the maximum separation. As will see later, this helps us to isolate the interface 
Fermi arc mediated current from the bulk current through the slabs. 
The Hamiltonian $H_t$ is obtained from $H_b$  by rotating it about the $z$ axis by an angle 
$\pi/2$  anticlockwise.  Both the slabs possess Fermi arc surface states localized at  their open 
surfaces at $z=0$. When the two slabs are tunnel coupled at $z=0$ via $H_c$, their interface
Fermi arcs states get hybridized which leads to a new set of Fermi arc states 
(which we call reconstructed Fermi arcs) which are also exponentially localized at the junction. 
The structure of the 
interface Fermi arcs depends on the content of the  tunneling matrix  in $H_c$ and the 
tunnel coupling strength $\kappa$. We note that the slabs are identical up to a rotation. 
So the system remains unchanged under the exchange $t \leftrightarrow b$. 
Imposing this symmetry on the tunneling matrix in $H_c$, we get
\cite{faruk2021farecon}
\begin{align}
H_c = \sum_{\bf k} c^{\dagger}_{t, 0}({\bf k})(u\sigma_x + i\sigma_z) c_{b, 0}({\bf k}) + H.c, 
\label{eq:tunnel}
\end{align}
where the parameter $u$ is real and is  assumed positive. When the tunnel coupling $\kappa$ 
is turned on, the Fermi arcs of the two WSM slabs reconstruct at the junction. As explained in detail in the supplemental of Ref.\cite{chau2023magnetic}, to connect Weyl nodes from the top and bottom slabs, the 
reconstructed Fermi arc states  need to join the  projections of same chirality (homochiral 
connectivity). For a fixed $u=0.5$,  the reconstructed Fermi arc states  
are shown in Fig. \ref{fig:reconfa} for a series of values $\kappa$. Here, we wish to 
point out that
we have two different regimes for $u$:  (i) $u\le 1$ where the interface localized states 
consists of only two isolated Fermi arcs, and (ii) $u >1$ where  additional isolated 
Fermi loops appear in the interface BZ as $\kappa$ is increased (see Appendix). In 
this work, our main focus is on the electronic transport from one slab to the other mediated by the interface 
localized Fermi  arcs. The  presence of additional Fermi loops only complicates the analysis without affecting our  main conclusions,  since they do not connect the Weyl nodes of the top and bottom surfaces.   Hence, in what 
follows, we mainly focus on  the electronic transport mediated by the interface states 
in the regime $u\le1$. We  briefly discuss  electronic transport in the other regime 
where interface BZ host additional isolated Fermi loops in the Appendix. 

\begin{figure}[tbph]
 \centering
 \includegraphics[width=1.0\linewidth]{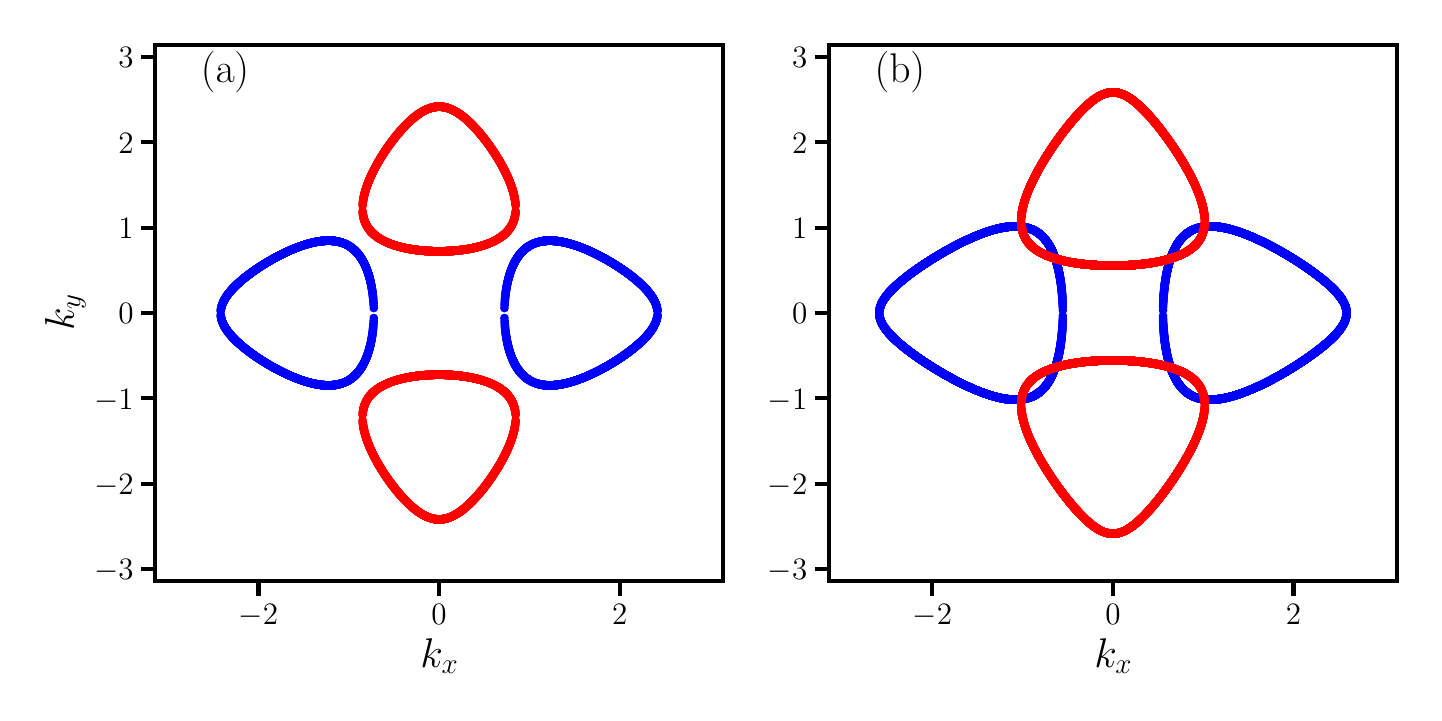}
 \caption{Figures show bulk Fermi surfaces around the Weyl 
 nodes for chemical potential (a) $\mu=1.5$ and (b) $\mu=1.7$. 
 The red contour belongs to the top slab and the blue contour is for the bottom slab. 
 The bulk Fermi surfaces start to overlap and contribute in the conduction 
 for  $\mu \ge \sqrt{8/3}$.  }
 \label{fig:fsovlp}
\end{figure}

\section{Tunnel Conductance in the absence of magnetic field }
\label{Sec:TCinAofMF}

In this section, we compute the conductance in the ${\hat z}$-direction in the absence of any applied magnetic field. 

Electron conduction from the bottom slab to the top slab  across the junction can be  
viewed as a transmission problem and the tunnel conductance across the junction can be 
obtained by the Landauer  formula 
\begin{align}
G(\mu) = \frac{e^2}{h} \sum_s M_s (\mu) T_s(\mu), 
\label{land}
\end{align}
where $\mu$ is the chemical potential, $M_s$ represents the number of modes 
of type $s$  and $T_s$ is their transmission probabilities.

For WSM slabs which are finite along the transverse directions  
($x$ and $y$-directions), we  have two type of modes:   (i) the bulk states near 
the Weyl nodes  and  (ii) the surface modes due to Fermi arcs states on the 
transverse surfaces.

Note that for the bulk states to be able to propagate from the bottom to the 
top slab, there must be a finite overlap between the bulk Fermi surfaces of the 
two WSM slabs \cite{tchoumakov201conservation, sousa2021gigantic, 
buccheri2022transport}.
However, this bulk contribution  can be excluded by taking a sufficiently 
small chemical potential so that this does not happen. This is illustrated 
in Fig. \ref{fig:fsovlp},
where we plot the bulk Fermi surfaces of the top and bottom slab for 
two different chemical potentials and show that there is no overlap in 
Fig. \ref{fig:fsovlp}(a), whereas there is overlap in Fig. \ref{fig:fsovlp}(b).

The fact that the interface Fermi arc states are localized at the junction, implies that they are 
naively  not expected to participate in electronic transport from one slab to the other 
across the junction. However,
since these states are extended across the interface, 
they can be instrumental in mediating transport via the transverse Fermi arc states.

\begin{figure}[tbph]
 \centering
 \includegraphics[width=1.0\linewidth]{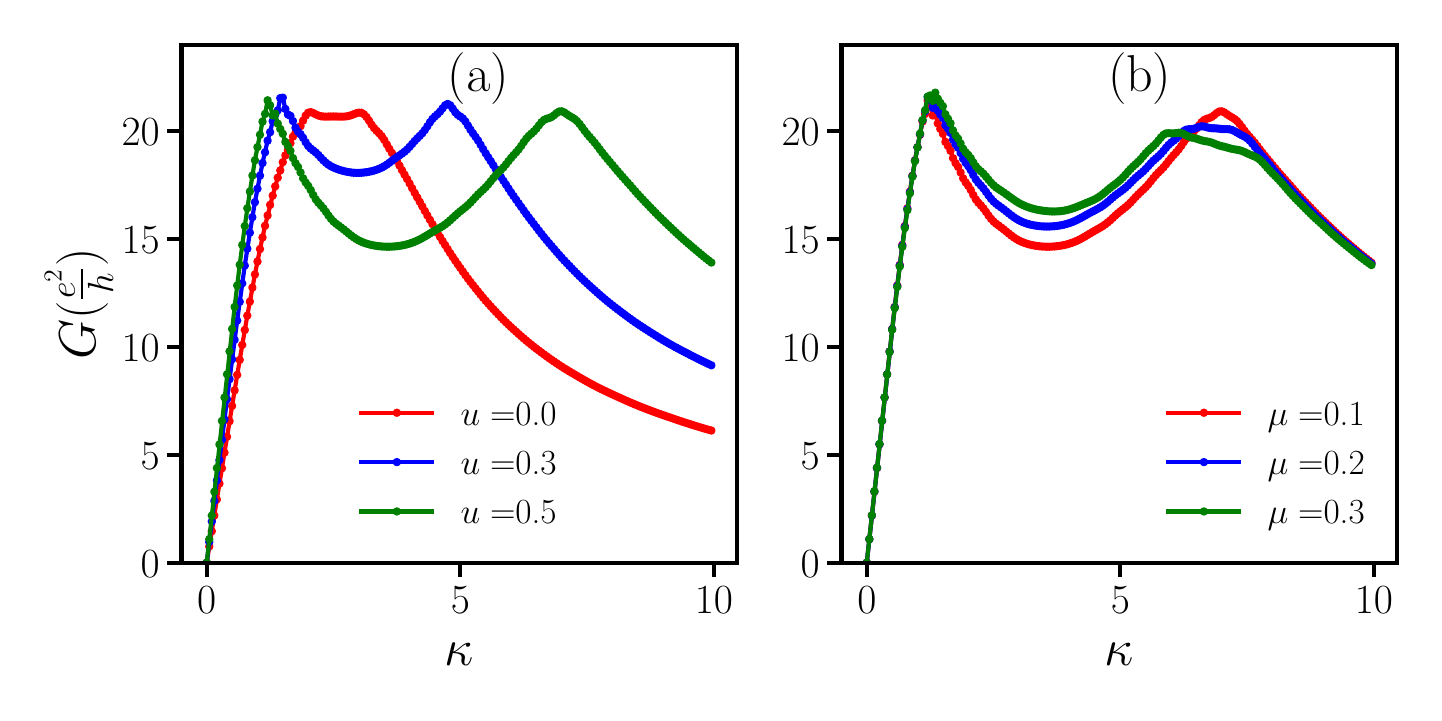}
 \caption{(a) The conductance is plotted for different values of $u$ 
 and a fixed $\mu=0.5$. Figure (b) shows conductance for a fixed $u = 0.5$ 
 and different values of the chemical potential $\mu$. We have 
 taken $L_x = L_y = L = 45$ in the units of the lattice constant and the Weyl 
 nodes' separation parameter $k_{0}=\pi/2$. We note that there are two peaks 
 in the tunnel conductance and the value of the $G_{peak} \approx \frac{e^2}{h} L/2$.
 The peaks correspond to the values of the tunnel coupling $\kappa$ at which 
 interface Fermi arcs almost straight. } 
 \label{fig:difv}
\end{figure}

Since the Weyl nodes in the bottom slab are separated along the $k_x$ axis by 
$2k_0$,  the $k_x$-$k_z$ surface BZ hosts (transverse) Fermi arc surface states which are 
localized on the $y=0$ and $y=L_y$ surfaces. At low energies, these states have 
dispersions $E(k_x, k_z) = k_z$ and  $E(k_x, k_z) = -k_z$,  so that  the electrons 
on these two surfaces move along the 
positive and negative $z$-directions respectively.  The top slab, on the other hand, 
has its Weyl nodes separated along the $k_y$-axis and its (transverse)
Fermi arc states localised along the $x=0$ and $x=L_x$ surface.
These states have dispersions $E(k_y, k_z) = k_z$ and  
$E(k_y, k_z) = -k_z$,  so that  the electrons on these two surfaces move along the 
positive and negative $z$-directions respectively.

Let us start with the upward moving electrons at $y=0$ on the bottom slab.
The picture is 
very simple when the bottom slab is not coupled to the top slab. Upon reaching the 
$z=0$ surface,  the electrons are taken over by the Fermi arc surfaces states living 
on the $z=0$ surface and hence propagate along the positive $y$-direction on 
the $z=0$ surface. After reaching  $y=L_y$, the electrons then move along the negative 
$z$-direction on the $y=L_y$ surface of the bottom slab.  So at zero tunnel coupling, 
the incident electrons occupying the transverse Fermi arc states of the $y=0$ surface  are fully reflected  
back to the bottom slab as transverse surface states on the $y=L_y$ surface. This reflection
of the surfaces states is mediated by the Fermi arc surface states localized at the $z=0$ 
surface.

In the presence of a finite tunnel coupling between the top and bottom slabs, at the interface,
the $z=0$  Fermi arc surface
states of the bottom slab hybridize with the adjacent 
Fermi arc 
surface states of the top slab. These new  reconstructed Fermi arc states are localized at the junction at 
$z=0$ and decay exponentially into the bulk of both the top and bottom slabs. However, as before, these interface Fermi arc states are extended in the $xy$ plane  of the junction. 
Now for a  finite tunnel coupling between the top and bottom slabs, the electrons which are 
traveling on the $y=0$ surface of the bottom slab,  have a finite  probability of transmission 
across the junction from the bottom to the  top slab. A transverse surface state along the $y=0$ surface moves upwards to the $z=0$ interface and via the reconstructed interface Fermi arc states at the junction, reaches the $x=0$ transverse state in the upper slab, thus contributing to the conductance.  We expect this conductance 
to depend mainly on the structure of the Fermi arcs at the interface BZ. Since the structure of the 
interface Fermi arcs evolves with tunnel coupling parameters  $\kappa$ and $u$ (see 
Fig. \ref{fig:reconfa}), we expect the tunnel conductance to  strongly depend on the values 
of  $\kappa$ and $u$. 
An important point to note is that because of the twist angle of $\pi/2$, due to which the transverse 
surface states are at right angles to each other (see Fig. \ref{fig:sideways}), there can be no
direct transfer of electrons from the bottom surface states to the top surface states - it can only occur 
via the interface states.

The total tunnel conductance is proportional to the number of modes, say $M_s$, in the 
transverse Fermi arc channel. For a Weyl node separation of $2k_0$, the total 
number of modes on transverse surfaces is proportional to the length of the 
Fermi arc on the $k_x$-$k_z$ surface BZ of the bottom slab: 
\begin{align}\label{Eq:ModesSurface}
    M_s = \frac{2k_0}{2\pi/L_x}. 
\end{align}
Substituting this in Eq.\ref{land}, 
for $k_0=\pi/2$, we note that  the maximum value  of the  tunnel conductance is $ G_{max} 
= \frac{e^2}{h}L_x/2$, which occurs when the transmission probability $T_s=1$
for some tunnel coupling $\kappa$. 
We will come back to this later.

Previous studies  \cite{buccheri2022transport} primarily focused on bulk electronic
transport through the junction of two WSM 
slabs twisted by a small angle, resulting in a finite overlap of their bulk Fermi surfaces. 
Although Ref. \cite{buccheri2022transport}  also examined electronic transport 
mediated by interface Fermi arc states, the discussion was largely qualitative and based 
on a continuum approximation applicable only for very small twist angles. 
In this work, 
we have chosen a large twist angle of $\pi/2$ between the slabs and perform a quantitative analysis 
of the tunnel conductance mediated by interface Fermi arc states. For such a large twist angle 
and a significant separation ($2k_0 = \pi$) between the Weyl nodes, the continuum 
approximation is no longer valid. Therefore, we employ a lattice model for the analysis. 
Unlike the continuum model, where substantial analytical progress can be made, 
calculating the tunnel conductance  in the full lattice model requires  numerical methods. 
We use KWANT \cite{groth2023kwant} simulations to numerically compute the tunnel 
conductance across 
the junction between the two WSM slabs. KWANT provides the total conductance, including 
contributions from both bulk states and Fermi arc surface states on the transverse 
surfaces. As previously mentioned, bulk states contribute only when there is a finite 
overlap between the bulk Fermi surfaces of the two WSM slabs.  We choose our chemical potential to ensure that there is no contribution from bulk states.
By selecting a relative 
twist angle of $\pi/2$ and a large separation ($2k_0 = \pi$) between the Weyl nodes, we ensure that there is a large range of chemical potential $\mu$ where this condition is met.

We compute the tunnel conductance across the junction as a function of  the tunnel coupling 
strength $\kappa$.  This is plotted in Fig. \ref{fig:difv}. The tunnel conductance is finite  
for nonzero values  of $\kappa$. For a fixed value of $u$ and  small values of $\kappa$
in the range $0\le \kappa \le \kappa_{u1}$, the conductance increases monotonically and  
reaches a peak value at $\kappa = \kappa_{u1}$. As $\kappa$ 
increases beyond $\kappa_{u1}$, the conductance first falls and then increases again 
to reach another peak at $\kappa=\kappa_{u2}$ before falling monotonically for 
large  values of $\kappa$. For $u=0.3$, the values of $\kappa_{u1}$ and $\kappa_{u2}$ 
are approximately 1.7 and 4.5 respectively. We notice that with increasing $u$, the value
of $\kappa_{u1}$ ($\kappa_{u2}$)  decreases (increases). 

This   behavior of the  tunnel conductance can be qualitatively understood as follows. 
At $\kappa \to 0$ and $\kappa \to \infty$, the interface behaves as an infinite potential 
barrier. The Fermi arc states of the two slabs at  $z=0$ are decoupled in this limits and they 
are orthogonal. As a result,  the incoming states get fully reflected and we find a vanishing 
tunnel conductance. This duality between $\kappa \to 0$ and $\kappa \to \infty$ has
been  noted earlier in Ref.\cite{faruk2021farecon}.

From the conductance plot in Fig. \ref{fig:difv}, we find that the values 
of tunnel conductance at peaks $G_{peak} \approx \frac{e^2}{h}L_x/2$ which is 
almost equal to the number of modes $M_s$ (see Eq. \ref{Eq:ModesSurface}) in 
the surface channel. This indicates that 
almost full transmission occurs when the tunnel coupling $\kappa$ is equal to 
$\kappa_{u1}$ or $\kappa_{u2}$ i.e. when the interface Fermi arcs are almost straight 
joining the projections of Weyl nodes of same chirality from the two slabs.

For  finite values of $\kappa$, the Fermi arc states of the two slabs at the junction  hybridize. 
So the incoming states  have a finite overlap with reconstructed interface Fermi arc states 
and can be transmitted to the top slab. 
When  $\kappa$ reaches $\kappa_{u1}$, the interface states 
joining the projection of the Weyl nodes  of the same chirality from the two slabs become 
straight, signifying that the velocities of a large number of states are parallel to each other. 
This prevents the different $k$ modes from scattering and ensures that a large fraction of the 
states from the bottom band is transmitted to the top band. As we increase  $\kappa$ 
beyond  $\kappa_{u1}$, the arc stops being a straight line and only again becomes a 
straight line at  $\kappa_{u2}$, where we see another peak in the conductance. 
Finally,  Fig.\ref{fig:difv}(b) clearly shows that the conductance does not vary as a 
function of the chemical potential for sufficiently small $\mu$ which is in accordance with 
the fact that the bulk states do not contribute to the conductance for small $\mu$.

\begin{figure}[tbph]
 \centering
 \includegraphics[width=0.8\linewidth]{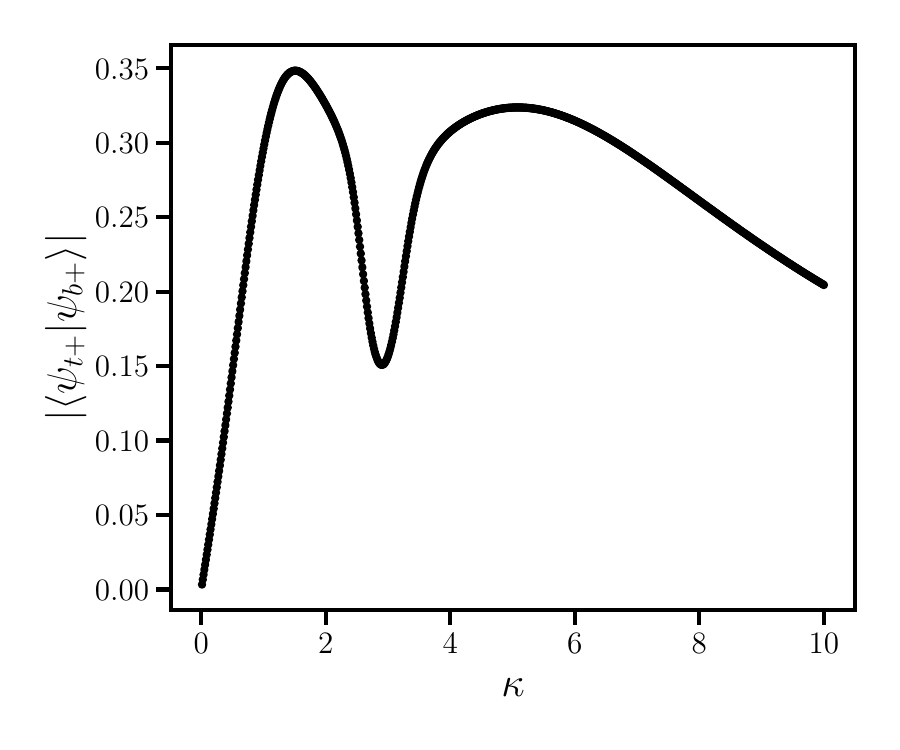}
 \caption{$|\langle \psi_{t+}| \psi_{b+} \rangle|$ is plotted with respect to $\kappa$. The 
 overlap is calculated at $\mu = 0.1$, $u = 0.5$. Here $\psi_{t+}$  and $\psi_{b+}$ are the 
 wavefunctions at momentum values close to the projections of the Weyl points of positive 
 chirality from the top and bottom slabs respectively.} 
 \label{fig:ovwf}
\end{figure}

A rough estimate of the hybridization between the Fermi arc states at the junction of the two slabs is 
given by an  overlap of the wavefunctions computed at two points (on the interface Fermi arcs)  which 
are close to the projections of Weyl points  of the same chirality from the two slabs. Let us denote 
$\psi_{t+}$  and $\psi_{b+}$  to be the wavefunctions at momentum values close to the projections 
of the Weyl points of positive chirality from the top and bottom slabs respectively. We compute 
$|\langle \psi_{t+}| \psi_{b+} \rangle|$ as a function of tunnel coupling and  plot it in Fig.\ref{fig:ovwf}.  
Note that  this overlap should go to  zero for $\kappa \to 0$ and $\kappa \to \infty$ because  
the Fermi arc states  of the two slabs localized at the junction  are orthogonal. We can see 
that the overlap has two peaks and the profile is qualitatively similar to the conductance plot in 
Fig. \ref{fig:difv} as expected.

\section{Magneto-Transport}

In this section, we study transport through the WSM junction in the presence of an external magnetic field perpendicular 
to the plane of the junction. This problem was initially  addressed 
in the case where the Fermi surfaces overlapped\cite{Kobayashi_2018, sousa2021gigantic}. In a later paper, Chaou {\it et al}\cite{chau2023magnetic} 
considered non-overlapping Fermi surfaces and 
showed that the electron transport in the presence of  longitudinal magnetic fields, gives rise to  a universal 
longitudinal magnetoconductance of $M_be^2/h$ for a sufficiently small fields, where  $M_b$ is the number of magnetic flux quanta through the interface. 
However, they did not include the contribution of the transverse Fermi arcs. Here, to put our work in context, we briefly
review their argument before we go on to include the contribution  of the transverse Fermi arcs.

\begin{figure}[tbph]
	\centering
	\includegraphics[width=0.7\linewidth, height=0.7\linewidth]{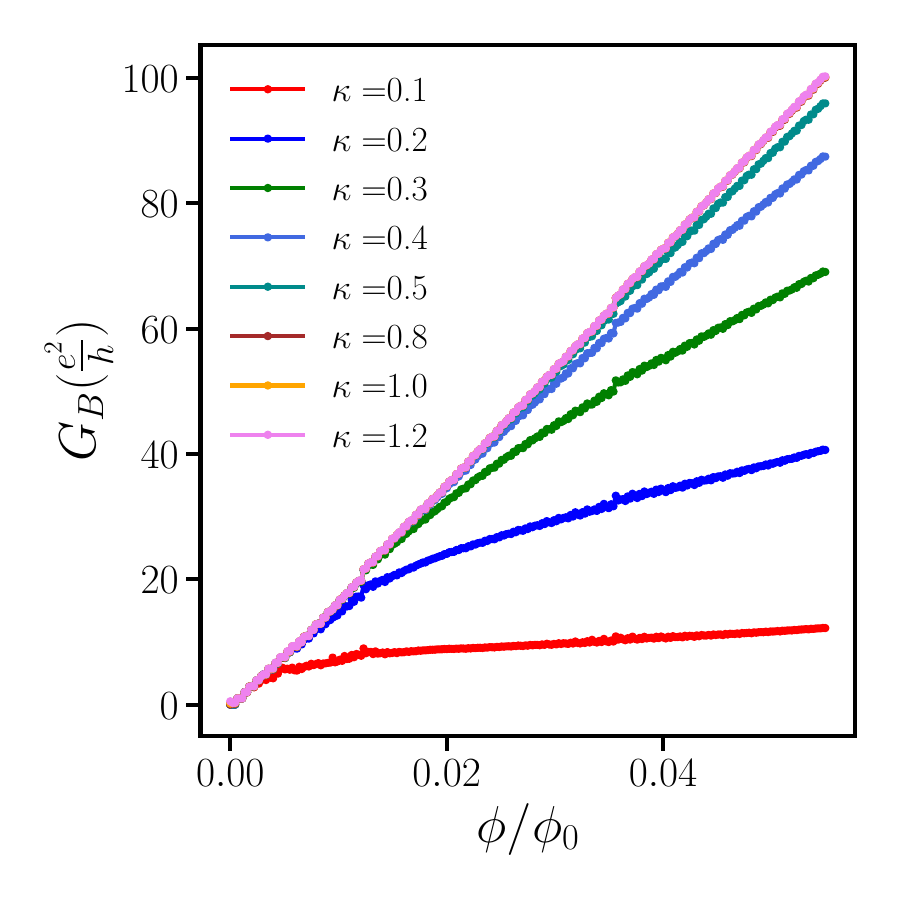}
 \caption{The tunnel conductance $G_B$ due to the bulk chiral Landau levels  in the presence of magnetic 
 fields perpendicular to the  interface as a function of the flux $\phi =
 Ba^2$ per unit cell  in  units of the  flux quantum $\phi_0 = h/e$, for various values of the tunneling 
 strength $\kappa$. The system is periodic in the $y$-direction and $L_x=L_y=45$ (in units of lattice 
 constant $a$).  We note that the  jaggedness in the plots are a numerical artifact.  }
 \label{fig:G_mag_periodic}
\end{figure}

When the Fermi surfaces of the two WSM slabs do not have overlap in the interface BZ, all but the chiral Landau levels 
are perfectly reflecting. The presence of a perpendicular magnetic field leads to $M_b$-fold degenerate chiral 
Landau levels at each Weyl 
node (where $M_b$ is the number of magnetic flux quanta through the interface). The positive chirality Weyl nodes 
have chiral modes  which disperse  along the positive $z$-direction  and 
the modes  which disperse along the negative $z$-direction are  associated with the negative chirality Weyl nodes. 
Electrons associated with the  appropriate Weyl node of positive chirality  travel along the positive $z$-direction 
in the bottom slab. After reaching the junction, 
they must continue to travel along the positive  $z$-direction. 
Since at  the junction, the  only available states are the interface Fermi arc states, the 
electrons  slide along the Fermi arc due to the Lorentz force
$-e {\bf v}_a \times {\bf B}$ from the projection of the positive 
chirality Weyl node of the bottom slab to the projection of 
the positive chirality Weyl node  of 
the top  slab. Thus electrons travel from  the bottom to  the top slab via the interface  localized 
Fermi arc states. The quantity ${\bf v}_a$, which is the group velocity of  the interface 
Fermi arc states, depends on ${\bf k}=(k_x, k_y)$  where  ${\bf k}$ is the momentum of the 
interface Fermi arc states. 

\begin{figure}[tbph]
	\centering
	\includegraphics[width=0.9\linewidth, height=0.8\linewidth]{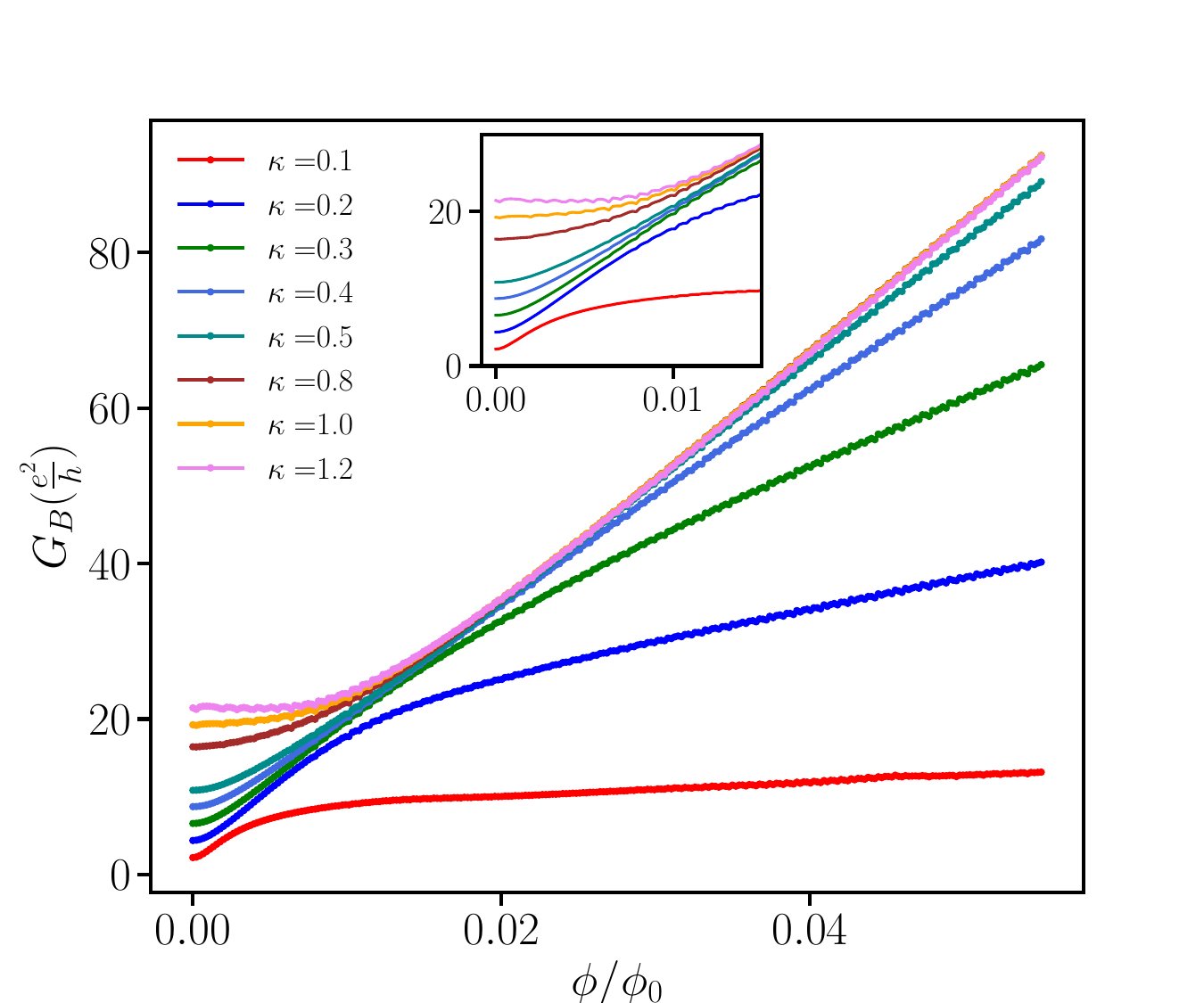}
 \caption{Total tunnel conductance $G_B$  (computed numerically)  in presence of 
 magnetic fields perpendicular to the  interface. The system is taken finite and 
 open along both the $x$ and $y$-directions with size same as in Fig. \ref{fig:G_mag_periodic}. 
 We note that conductance is finite at zero flux, because the Fermi arc surface states on 
 the transverse surfaces carry a finite current across the interface (see Fig. \ref{fig:difv}).  At 
 large fluxes, the conductance which is dominated by the bulk chiral Landau levels is 
 approximately described by the Eq. \ref{Eq:GB2}. The inset highlights the behaviour 
 of conductance for smaller fluxes, where surface contribution is significant. }
 \label{fig:G_mag_finite}
\end{figure}

Note  that we have two separate Fermi arcs in  the interface BZ. Besides the Fermi arc states joining the projections of the positive chirality Weyl nodes, we also have the Fermi arc states 
joining  the projection  of the negative  chirality Weyl nodes. These  act as a channel for 
transport of electrons from  the top to the bottom slab.  
There will be a net current 
flowing from the bottom to the top slab only when two slabs are subjected to a finite 
bias along the positive $z$-direction.

To compute the tunnel conductance between  the two slabs  along the $z$-direction, we again employ  
KWANT simulations. Unlike the zero field case, here both the bulk states (the chiral Landau levels)
and the transverse Fermi arc surface states carry current across the 
junction from the bottom to the top slab. However, it is possible to
isolate the two contributions by making our 
system periodic in either of  the transverse directions ($x$  or $y$), so 
that one of the transverse Fermi arcs, either from the bottom slab or the 
top slab drop out. However, we can not take the system 
to be periodic both along the  $x$ and $y$-directions,  because a uniform magnetic field 
along the $z$-direction will  break translations symmetry in at least  one of the two  transverse 
directions. However, since the bottom and top  WSM slabs
have  surface states in perpendicular directions, even making one transverse direction periodic is 
sufficient to remove the transverse Fermi arc contributions.

\begin{figure}[tbph]
	\centering
	\includegraphics[width=0.7\linewidth]{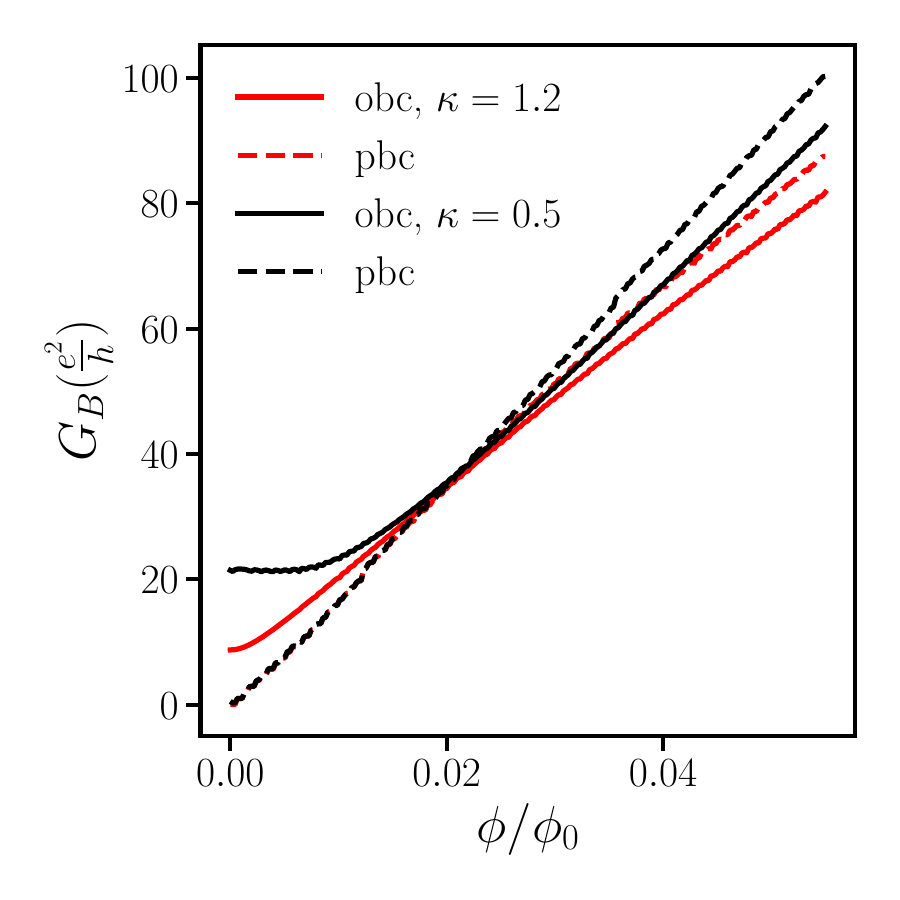}
 \caption{Magnetocoductance for periodic boundary conditions (no transverse surface states)  is compared with 
 that for  open boundary conditions. The system size is the same as in Fig. \ref{fig:G_mag_finite}. 
 Solid (dashed) lines represent conductance due to the bulk states (bulk + surface states). The 
 total conductance  due to the bulk+surface  becomes smaller than the conductance 
 due to only the bulk states after a value of flux $\phi_c/\phi_0 \approx 0.02$ which 
 corresponds to a value of field $B_c \sim 40$T for $L_x=L_y=45a$, $a=1$nm, and Weyl node 
 separation $2k_0=\pi$. }
 \label{fig:G_mag_diff}
\end{figure}

In this limit, we reproduce the results (see Fig. \ref{fig:G_mag_periodic}) in Ref.\cite{chau2023magnetic} 
and we note that the conductance is linear in the magnetic field for small  fields
and is  given by 
\begin{align}\label{Eq:GB1}
G_B  = \frac{e^2}{h} M_b T , 
\end{align}
where $M_b \propto B$ is the degeneracy (or number of modes) of  the chiral Landau level 
and $T=1$. This remains true (i.e. $T=1$) as long as the two oppositely 
dispersing chiral Landau levels remain 
decoupled in the bulk, which is only satisfied when the  magnetic length 
$l_B^{-1} \gg 2k_0$ ($2k_0$ is the momentum space separation of two Weyl of 
opposite chiralities)\cite{Kim_breakdown_2017, Chan_emergence_2017, Abdulla_timerev_2022, Abdulla_pairwise_2024}. 
However, there is reconstruction of the Fermi arcs at the interface BZ, for non-zero $\kappa$. 
Due to this, the two Fermi arcs can be coupled if the  separation $\Delta$ between the Fermi arcs  
is of the order $\sim l_B^{-1}$.  This can cause reflection at the interface and reduce
the transmission from unity. As shown in Ref.\cite{chau2023magnetic}, the conductance in this case is given by 
\begin{align}\label{Eq:GB2}
G_B  = \frac{e^2}{h}  M_b \left(1 - e^{-B_0/B}  \right). 
\end{align}
where $B_0 = \frac{\pi}{4} \Delta^2 \tan(\theta/2)$. 
Here $\theta$ is the angle at which 
the Fermi arcs of the two slabs crosses in the interface BZ at zero coupling. In our model 
$\theta=\pi/2$, so $B_0 =\frac{\pi}{4} \Delta^2$.

Given our understanding of electronic transport via the bulk chiral Landau levels, 
we can now examine the overall magnetoconductance including the contribution of the 
transverse
Fermi arc surface states. The numerical results are presented in Fig. \ref{fig:G_mag_finite}. 
At high fluxes, the total conductance remains nearly the same as in the previous 
scenario, where only the bulk chiral Landau levels contributed to the conductance. 
An additional contribution to the conductance is observed only at low fields, where we 
find that the Fermi arc states on the transverse surfaces significantly impact the 
conductance.  The reason 
is the following: For a fixed tunnel coupling strength $\kappa$ and a given channel, 
the conductance  depends on the number of modes in the channel and their 
transmission probabilities. Here, a crucial factor to consider is the number of modes 
in the bulk and surface channels. The number of modes in the bulk channel is given 
by  the degeneracy $M_b = L_x L_y B/\phi_0$ of the chiral Landau level which 
increases linearly with $B$ and is also proportional to the area of the sample.  
On  the other hand, the number of modes in the surface channel  $M_s = 2k_0 L_x/2\pi$ which is  
independent of the value of magnetic fields and is also only linearly dependent on the size of the sample. The number of bulk and surface modes becomes comparable when flux 
$\phi_c/\phi_0 \sim k_0 a^2/L_y$ 
($a$ is lattice constant). Since for a macroscopic system $k_0 a\ll L_y/a$, 
clearly the critical value of of flux $\phi_c$ below which the transverse  Fermi arcs 
play a role is small. Above this flux, 
it is the bulk states (chiral Landau levels) which dominate the conductance and hence
the conductance is approximately given by the Eq. \ref{Eq:GB2}.  
However, for a mesoscopic system with, say, $L_x= L_y = 10$ micron and a lattice 
constant $a=1$ A$^0$, the typical value of  magnetic field $B_c \sim 2$ T, which is notably high. 
Consequently, the surface contribution to the tunnel conductance cannot be ignored; 
in fact, it is the surface contribution that primarily influences the electronic transport 
under any experimentally relevant fields for mesoscopic or smaller samples.

In Fig.\ref{fig:G_mag_diff}, we emphasize the difference to the conductance 
at low flux ($\phi<\phi_c$) due to the contribution of the transverse 
Fermi arcs. We note that the conductance is finite even at 
zero magnetic field (as would be expected from our results in Sec. III). But 
at larger magnetic fields, 
the conductance is dominated by the bulk chiral levels and scales linearly with 
magnetic field as expected from Eq. \ref{Eq:GB2} (before it reaches the saturation value).
In the presence of both bulk and surface channels, one might naively  expect the total 
conductance to be always higher than the periodic case where  only bulk chiral 
channels contribute to  transport across the interface. This expectation holds 
for flux $\phi<\phi_c$, where the number of surface modes is significantly 
higher than the number of bulk chiral modes. 
However, at slightly higher flux $\phi>\phi_c$, where their contribution to 
the overall conductance is much smaller, the situation can change 
(see Fig. \ref{fig:G_mag_diff}). This could be  due to mixing between  the transverse Fermi 
arcs and the bulk chiral states.
We note that for low magnetic fields, in the vicinity of the Weyl nodes, the transverse surface states are not 
completely isolated from the 
bulk chiral states- there is a significant  mixing between them.
This can introduce additional scatterings which can complicate the electron transport from 
one slab to other at low fields. An understanding of how the conductance
scales with magnetic  fields in the low field regime where the bulk and surface 
modes are comparable requires a detail and more involved analytical study which is 
beyond the scope of our numerical study.

\section{Discussion and Conclusion}

We have studied magnetotransport across twisted  Weyl semimetals including 
transport due to the transverse surface Fermi arcs, which was ignored in 
earlier studies. We find that this gives rise to transport 
via the interface Fermi arcs even in the absence of magnetic fields. The conductance 
is proportional to the linear dimension, in the direction perpendicular to the direction 
of separation of the Weyl nodes  of the sample. For large twist angles or when 
the projections of the Weyl nodes from the two slabs on  the interface BZ are 
well separated, this is the only contribution to the tunnel conductance.

In the presence of magnetic fields perpendicular to the junction, transverse 
Fermi arc states as well as the bulk states (chiral Landau levels) contribute 
to the tunnel conductance. The bulk contribution 
to the tunnel conductance is proportional to the area of the sample. For 
macroscopic systems,  the surface contribution (which is proportional to the 
linear dimension of the sample) is much smaller 
than the bulk contribution and maybe neglected. However, for a mesoscopic or smaller
systems, the surface contribution dominates the tunnel conductance for any 
experimentally relevant magnetic fields. 

The conductance has been computed using a Landauer-Buttiker type numerical  approach
for a twist angle of $\theta = \pi/2$, which ensured that the lattice sites of the bottom and top  slabs were  aligned at the interface. This alignment offered a significant computational advantage over  the non-aligned cases, which would require  superlattices featuring  much larger unit cells.
However,  we note that the conductance is not dependent on the twist angle, but on the number
of modes and the transmission.
Hence, we believe that our results are generic and 
will be true for other  twist angles and Weyl node separations 
provided that the bulk Fermi surfaces of the two slabs have no
overlap in the interface Brillouin zone and
also provided that there is a sufficient region in momentum space, where 
the Fermi arcs are parallel, and there is low backscattering
between the modes.


\begin{acknowledgements}
The authors would like thank Ganpathy Murthy for valuable discussions. NB would like to
thank the Infosys Foundation for financial support. NB and 
FA acknowledge the  International Centre for Theoretical Sciences (ICTS), where 
a  significant part of this work was done, for their hospitality, funding, and 
kind support toward academic collaboration. 
\end{acknowledgements}


\appendix

\section{Reconstructed Fermi arcs and tunnel conductance for $u>1$ }

\begin{figure}
	\centering
	\includegraphics[width=1.0\linewidth]{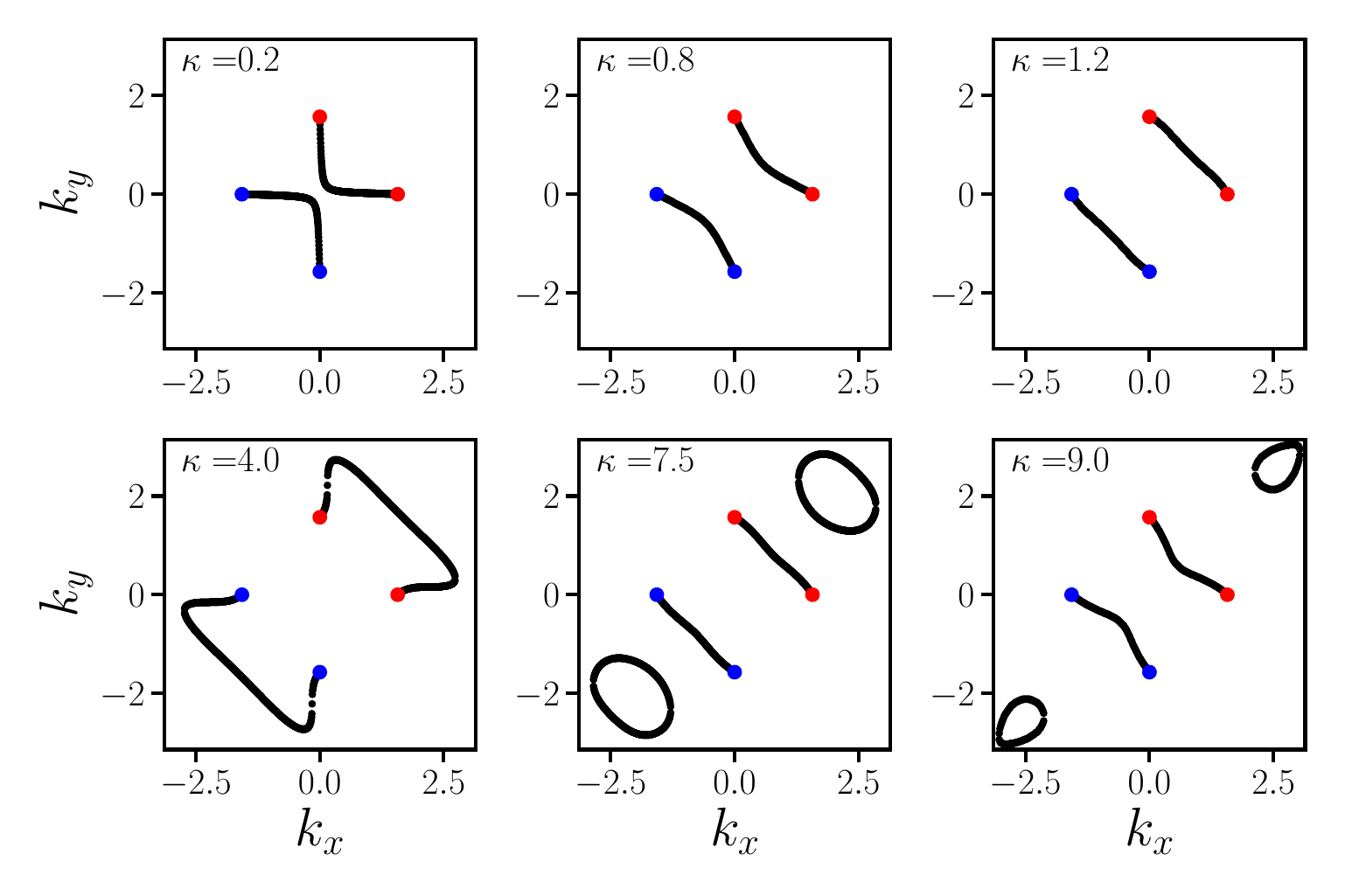}
 \caption{Interface Fermi arc plots in the interface BZ are 
 presented for several values of the tunnel coupling, $\kappa$, with 
 all plots using $u = 2$. The reconstructed Fermi arcs appear nearly 
 straight at $\kappa \approx \kappa_s$. For $\kappa$ values below $\kappa_s$, 
 the Fermi arcs curve inward, while increasing $\kappa$ beyond $\kappa_s$ 
 causes the arcs to curve outward. Near $\kappa \approx 6.7$, each of the 
 two Fermi arcs splits, leading to the formation of a pair of isolated 
 Fermi loops. }
 \label{fig:reconfa_new}
\end{figure}

In the main text we discussed the reconstruction and the transport for 
the tunneling parameter $u<1$. Here, we generalise it to the case 
when $u>1$. While the conductance  shows no change in qualitative 
behavious  as a function of $u$,  the Fermi arc structure does show some 
interesting change of features for $u>1$. For a fixed $u = 2.0$, the 
reconstructed Fermi arc states are shown in Fig. \ref{fig:reconfa_new} 
for a series of values of $\kappa$. The main contrasting feature from  
the $u<1$ case is the existence of an additional Fermi loop states for 
certain values of $\kappa$. As $\kappa$ is increased beyond its value 
where the Fermi arcs are almost straight, the Fermi arcs develop 
opposite curvature. When tunnel coupling is above a certain value, 
say $\kappa_c$, which depends on the value of $u$, the Fermi arc 
splits and form an isolated  loop. As $u \to 1$, the value of 
$\kappa_c$ at which Fermi loop appears approaches to infinity.

\begin{figure}
	\centering
	\includegraphics[width=1.0\linewidth]{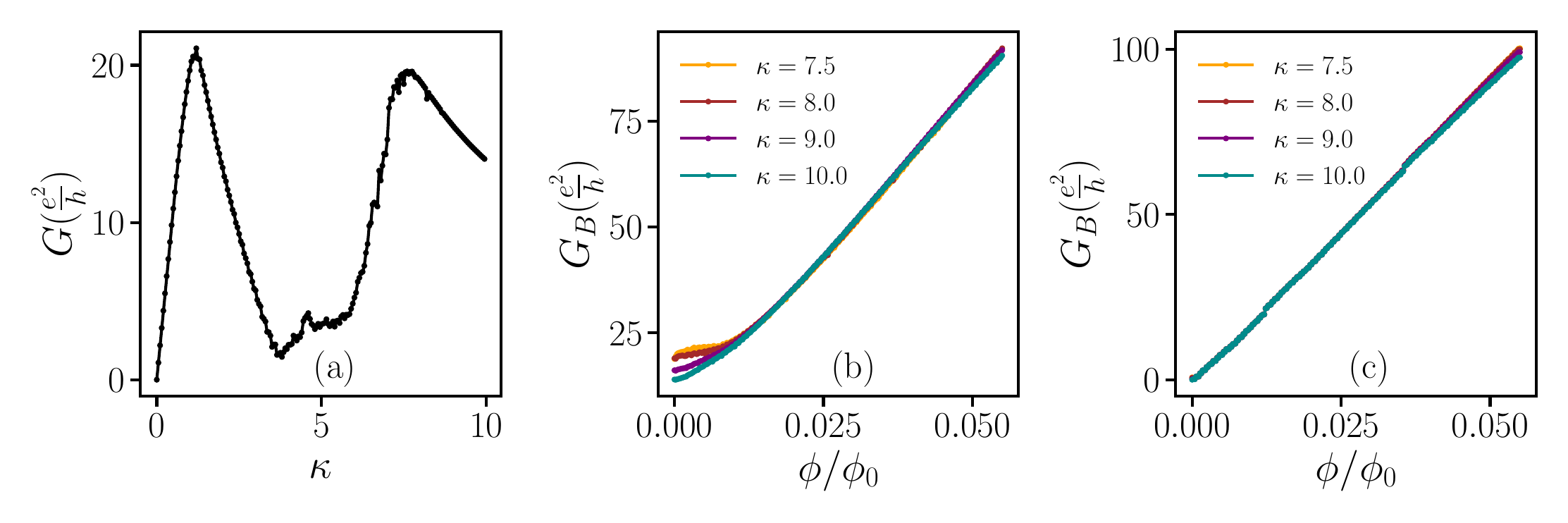}
 \caption{Figure (a) illustrates tunnel conductance as a function of 
 the tunnel coupling, $\kappa$, at zero magnetic field. Figure (b) displays 
 conductance as a function of the applied flux through the interface, with 
 open boundary conditions in the transverse directions. Figure (c) shows 
 conductance as a function of flux in the scenario where no transverse 
 surface arc states are present, meaning only the bulk chiral Landau levels 
 contribute to the tunnel conductance. For all the cases $u=2.0$ and 
 $L_x=L_y=45$ in units of lattice constant $a$. }
 \label{fig:finite_new_womag}
\end{figure}

We compute the tunnel conductance across the junction as a function of the tunnel coupling strength 
$\kappa$ and this is plotted in Fig. \ref{fig:finite_new_womag}a. We find that the  conductance 
is finite for finite values of $\kappa$ and the features are similar to what was observed for the 
conductance for $u<1$. In presence of magnetic fields perpendicular to the junction, the conductance as a function of applied flux  is shown in Fig. \ref{fig:finite_new_womag}b for a few $\kappa$ values  near the $\kappa_c$ where additional Fermi loops exist. Here also we 
find that the features are qualitatively  similar to what was observed for the 
conductance for $u<1$.

\bibliography{main}

\begin{thebibliography}{46}%
\makeatletter
\providecommand \@ifxundefined [1]{%
 \@ifx{#1\undefined}
}%
\providecommand \@ifnum [1]{%
 \ifnum #1\expandafter \@firstoftwo
 \else \expandafter \@secondoftwo
 \fi
}%
\providecommand \@ifx [1]{%
 \ifx #1\expandafter \@firstoftwo
 \else \expandafter \@secondoftwo
 \fi
}%
\providecommand \natexlab [1]{#1}%
\providecommand \enquote  [1]{``#1''}%
\providecommand \bibnamefont  [1]{#1}%
\providecommand \bibfnamefont [1]{#1}%
\providecommand \citenamefont [1]{#1}%
\providecommand \href@noop [0]{\@secondoftwo}%
\providecommand \href [0]{\begingroup \@sanitize@url \@href}%
\providecommand \@href[1]{\@@startlink{#1}\@@href}%
\providecommand \@@href[1]{\endgroup#1\@@endlink}%
\providecommand \@sanitize@url [0]{\catcode `\\12\catcode `\$12\catcode
  `\&12\catcode `\#12\catcode `\^12\catcode `\_12\catcode `\%12\relax}%
\providecommand \@@startlink[1]{}%
\providecommand \@@endlink[0]{}%
\providecommand \url  [0]{\begingroup\@sanitize@url \@url }%
\providecommand \@url [1]{\endgroup\@href {#1}{\urlprefix }}%
\providecommand \urlprefix  [0]{URL }%
\providecommand \Eprint [0]{\href }%
\providecommand \doibase [0]{https://doi.org/}%
\providecommand \selectlanguage [0]{\@gobble}%
\providecommand \bibinfo  [0]{\@secondoftwo}%
\providecommand \bibfield  [0]{\@secondoftwo}%
\providecommand \translation [1]{[#1]}%
\providecommand \BibitemOpen [0]{}%
\providecommand \bibitemStop [0]{}%
\providecommand \bibitemNoStop [0]{.\EOS\space}%
\providecommand \EOS [0]{\spacefactor3000\relax}%
\providecommand \BibitemShut  [1]{\csname bibitem#1\endcsname}%
\let\auto@bib@innerbib\@empty
\bibitem [{\citenamefont {Murakami}(2007)}]{Murakami_2007}%
  \BibitemOpen
  \bibfield  {author} {\bibinfo {author} {\bibfnamefont {S.}~\bibnamefont
  {Murakami}},\ }\bibfield  {title} {\bibinfo {title} {Phase transition between
  the quantum spin hall and insulator phases in 3d: emergence of a topological
  gapless phase},\ }\href {https://doi.org/10.1088/1367-2630/9/9/356}
  {\bibfield  {journal} {\bibinfo  {journal} {New Journal of Physics}\ }\textbf
  {\bibinfo {volume} {9}},\ \bibinfo {pages} {356} (\bibinfo {year}
  {2007})}\BibitemShut {NoStop}%
\bibitem [{\citenamefont {Wan}\ \emph {et~al.}(2011)\citenamefont {Wan},
  \citenamefont {Turner}, \citenamefont {Vishwanath},\ and\ \citenamefont
  {Savrasov}}]{Wan_Savrasov_2011}%
  \BibitemOpen
  \bibfield  {author} {\bibinfo {author} {\bibfnamefont {X.}~\bibnamefont
  {Wan}}, \bibinfo {author} {\bibfnamefont {A.~M.}\ \bibnamefont {Turner}},
  \bibinfo {author} {\bibfnamefont {A.}~\bibnamefont {Vishwanath}},\ and\
  \bibinfo {author} {\bibfnamefont {S.~Y.}\ \bibnamefont {Savrasov}},\
  }\bibfield  {title} {\bibinfo {title} {Topological semimetal and fermi-arc
  surface states in the electronic structure of pyrochlore iridates},\ }\href
  {https://doi.org/10.1103/PhysRevB.83.205101} {\bibfield  {journal} {\bibinfo
  {journal} {Phys. Rev. B}\ }\textbf {\bibinfo {volume} {83}},\ \bibinfo
  {pages} {205101} (\bibinfo {year} {2011})}\BibitemShut {NoStop}%
\bibitem [{\citenamefont {Yang}\ \emph {et~al.}(2011)\citenamefont {Yang},
  \citenamefont {Lu},\ and\ \citenamefont {Ran}}]{Yang_Ran_2011}%
  \BibitemOpen
  \bibfield  {author} {\bibinfo {author} {\bibfnamefont {K.-Y.}\ \bibnamefont
  {Yang}}, \bibinfo {author} {\bibfnamefont {Y.-M.}\ \bibnamefont {Lu}},\ and\
  \bibinfo {author} {\bibfnamefont {Y.}~\bibnamefont {Ran}},\ }\bibfield
  {title} {\bibinfo {title} {Quantum hall effects in a weyl semimetal: Possible
  application in pyrochlore iridates},\ }\href
  {https://doi.org/10.1103/PhysRevB.84.075129} {\bibfield  {journal} {\bibinfo
  {journal} {Phys. Rev. B}\ }\textbf {\bibinfo {volume} {84}},\ \bibinfo
  {pages} {075129} (\bibinfo {year} {2011})}\BibitemShut {NoStop}%
\bibitem [{\citenamefont {Burkov}\ and\ \citenamefont
  {Balents}(2011)}]{Burkov_Balents_2011}%
  \BibitemOpen
  \bibfield  {author} {\bibinfo {author} {\bibfnamefont {A.~A.}\ \bibnamefont
  {Burkov}}\ and\ \bibinfo {author} {\bibfnamefont {L.}~\bibnamefont
  {Balents}},\ }\bibfield  {title} {\bibinfo {title} {Weyl semimetal in a
  topological insulator multilayer},\ }\href
  {https://doi.org/10.1103/PhysRevLett.107.127205} {\bibfield  {journal}
  {\bibinfo  {journal} {Phys. Rev. Lett.}\ }\textbf {\bibinfo {volume} {107}},\
  \bibinfo {pages} {127205} (\bibinfo {year} {2011})}\BibitemShut {NoStop}%
\bibitem [{\citenamefont {Xu}\ \emph {et~al.}(2011)\citenamefont {Xu},
  \citenamefont {Weng}, \citenamefont {Wang}, \citenamefont {Dai},\ and\
  \citenamefont {Fang}}]{Xu_Fang_2011}%
  \BibitemOpen
  \bibfield  {author} {\bibinfo {author} {\bibfnamefont {G.}~\bibnamefont
  {Xu}}, \bibinfo {author} {\bibfnamefont {H.}~\bibnamefont {Weng}}, \bibinfo
  {author} {\bibfnamefont {Z.}~\bibnamefont {Wang}}, \bibinfo {author}
  {\bibfnamefont {X.}~\bibnamefont {Dai}},\ and\ \bibinfo {author}
  {\bibfnamefont {Z.}~\bibnamefont {Fang}},\ }\bibfield  {title} {\bibinfo
  {title} {Chern semimetal and the quantized anomalous hall effect in
  ${\mathrm{hgcr}}_{2}{\mathrm{se}}_{4}$},\ }\href
  {https://doi.org/10.1103/PhysRevLett.107.186806} {\bibfield  {journal}
  {\bibinfo  {journal} {Phys. Rev. Lett.}\ }\textbf {\bibinfo {volume} {107}},\
  \bibinfo {pages} {186806} (\bibinfo {year} {2011})}\BibitemShut {NoStop}%
\bibitem [{\citenamefont {Lv}\ \emph {et~al.}(2015{\natexlab{a}})\citenamefont
  {Lv}, \citenamefont {Weng}, \citenamefont {Fu}, \citenamefont {Wang},
  \citenamefont {Miao}, \citenamefont {Ma}, \citenamefont {Richard},
  \citenamefont {Huang}, \citenamefont {Zhao}, \citenamefont {Chen},
  \citenamefont {Fang}, \citenamefont {Dai}, \citenamefont {Qian},\ and\
  \citenamefont {Ding}}]{Lv_Ding_2015a}%
  \BibitemOpen
  \bibfield  {author} {\bibinfo {author} {\bibfnamefont {B.~Q.}\ \bibnamefont
  {Lv}}, \bibinfo {author} {\bibfnamefont {H.~M.}\ \bibnamefont {Weng}},
  \bibinfo {author} {\bibfnamefont {B.~B.}\ \bibnamefont {Fu}}, \bibinfo
  {author} {\bibfnamefont {X.~P.}\ \bibnamefont {Wang}}, \bibinfo {author}
  {\bibfnamefont {H.}~\bibnamefont {Miao}}, \bibinfo {author} {\bibfnamefont
  {J.}~\bibnamefont {Ma}}, \bibinfo {author} {\bibfnamefont {P.}~\bibnamefont
  {Richard}}, \bibinfo {author} {\bibfnamefont {X.~C.}\ \bibnamefont {Huang}},
  \bibinfo {author} {\bibfnamefont {L.~X.}\ \bibnamefont {Zhao}}, \bibinfo
  {author} {\bibfnamefont {G.~F.}\ \bibnamefont {Chen}}, \bibinfo {author}
  {\bibfnamefont {Z.}~\bibnamefont {Fang}}, \bibinfo {author} {\bibfnamefont
  {X.}~\bibnamefont {Dai}}, \bibinfo {author} {\bibfnamefont {T.}~\bibnamefont
  {Qian}},\ and\ \bibinfo {author} {\bibfnamefont {H.}~\bibnamefont {Ding}},\
  }\bibfield  {title} {\bibinfo {title} {Experimental discovery of weyl
  semimetal taas},\ }\href {https://doi.org/10.1103/PhysRevX.5.031013}
  {\bibfield  {journal} {\bibinfo  {journal} {Phys. Rev. X}\ }\textbf {\bibinfo
  {volume} {5}},\ \bibinfo {pages} {031013} (\bibinfo {year}
  {2015}{\natexlab{a}})}\BibitemShut {NoStop}%
\bibitem [{\citenamefont {Lv}\ \emph {et~al.}(2015{\natexlab{b}})\citenamefont
  {Lv}, \citenamefont {Xu}, \citenamefont {Weng}, \citenamefont {Ma},
  \citenamefont {Richard}, \citenamefont {Huang}, \citenamefont {Zhao},
  \citenamefont {Chen}, \citenamefont {Matt}, \citenamefont {Bisti},
  \citenamefont {Strocov}, \citenamefont {Mesot}, \citenamefont {Fang},
  \citenamefont {Dai}, \citenamefont {Qian}, \citenamefont {Shi},\ and\
  \citenamefont {Ding}}]{Lv_Ding_2015b}%
  \BibitemOpen
  \bibfield  {author} {\bibinfo {author} {\bibfnamefont {B.~Q.}\ \bibnamefont
  {Lv}}, \bibinfo {author} {\bibfnamefont {N.}~\bibnamefont {Xu}}, \bibinfo
  {author} {\bibfnamefont {H.~M.}\ \bibnamefont {Weng}}, \bibinfo {author}
  {\bibfnamefont {J.~Z.}\ \bibnamefont {Ma}}, \bibinfo {author} {\bibfnamefont
  {P.}~\bibnamefont {Richard}}, \bibinfo {author} {\bibfnamefont {X.~C.}\
  \bibnamefont {Huang}}, \bibinfo {author} {\bibfnamefont {L.~X.}\ \bibnamefont
  {Zhao}}, \bibinfo {author} {\bibfnamefont {G.~F.}\ \bibnamefont {Chen}},
  \bibinfo {author} {\bibfnamefont {C.~E.}\ \bibnamefont {Matt}}, \bibinfo
  {author} {\bibfnamefont {F.}~\bibnamefont {Bisti}}, \bibinfo {author}
  {\bibfnamefont {V.~N.}\ \bibnamefont {Strocov}}, \bibinfo {author}
  {\bibfnamefont {J.}~\bibnamefont {Mesot}}, \bibinfo {author} {\bibfnamefont
  {Z.}~\bibnamefont {Fang}}, \bibinfo {author} {\bibfnamefont {X.}~\bibnamefont
  {Dai}}, \bibinfo {author} {\bibfnamefont {T.}~\bibnamefont {Qian}}, \bibinfo
  {author} {\bibfnamefont {M.}~\bibnamefont {Shi}},\ and\ \bibinfo {author}
  {\bibfnamefont {H.}~\bibnamefont {Ding}},\ }\bibfield  {title} {\bibinfo
  {title} {Observation of weyl nodes in taas},\ }\href
  {https://doi.org/10.1038/nphys3426} {\bibfield  {journal} {\bibinfo
  {journal} {Nature Physics}\ }\textbf {\bibinfo {volume} {11}},\ \bibinfo
  {pages} {724} (\bibinfo {year} {2015}{\natexlab{b}})}\BibitemShut {NoStop}%
\bibitem [{\citenamefont {Xu}\ \emph {et~al.}(2015{\natexlab{a}})\citenamefont
  {Xu}, \citenamefont {Belopolski}, \citenamefont {Alidoust}, \citenamefont
  {Neupane}, \citenamefont {Bian}, \citenamefont {Zhang}, \citenamefont
  {Sankar}, \citenamefont {Chang}, \citenamefont {Yuan}, \citenamefont {Lee},
  \citenamefont {Huang}, \citenamefont {Zheng}, \citenamefont {Ma},
  \citenamefont {Sanchez}, \citenamefont {Wang}, \citenamefont {Bansil},
  \citenamefont {Chou}, \citenamefont {Shibayev}, \citenamefont {Lin},
  \citenamefont {Jia},\ and\ \citenamefont {Hasan}}]{Xu_Hasan_2015a}%
  \BibitemOpen
  \bibfield  {author} {\bibinfo {author} {\bibfnamefont {S.-Y.}\ \bibnamefont
  {Xu}}, \bibinfo {author} {\bibfnamefont {I.}~\bibnamefont {Belopolski}},
  \bibinfo {author} {\bibfnamefont {N.}~\bibnamefont {Alidoust}}, \bibinfo
  {author} {\bibfnamefont {M.}~\bibnamefont {Neupane}}, \bibinfo {author}
  {\bibfnamefont {G.}~\bibnamefont {Bian}}, \bibinfo {author} {\bibfnamefont
  {C.}~\bibnamefont {Zhang}}, \bibinfo {author} {\bibfnamefont
  {R.}~\bibnamefont {Sankar}}, \bibinfo {author} {\bibfnamefont
  {G.}~\bibnamefont {Chang}}, \bibinfo {author} {\bibfnamefont
  {Z.}~\bibnamefont {Yuan}}, \bibinfo {author} {\bibfnamefont {C.-C.}\
  \bibnamefont {Lee}}, \bibinfo {author} {\bibfnamefont {S.-M.}\ \bibnamefont
  {Huang}}, \bibinfo {author} {\bibfnamefont {H.}~\bibnamefont {Zheng}},
  \bibinfo {author} {\bibfnamefont {J.}~\bibnamefont {Ma}}, \bibinfo {author}
  {\bibfnamefont {D.~S.}\ \bibnamefont {Sanchez}}, \bibinfo {author}
  {\bibfnamefont {B.}~\bibnamefont {Wang}}, \bibinfo {author} {\bibfnamefont
  {A.}~\bibnamefont {Bansil}}, \bibinfo {author} {\bibfnamefont
  {F.}~\bibnamefont {Chou}}, \bibinfo {author} {\bibfnamefont {P.~P.}\
  \bibnamefont {Shibayev}}, \bibinfo {author} {\bibfnamefont {H.}~\bibnamefont
  {Lin}}, \bibinfo {author} {\bibfnamefont {S.}~\bibnamefont {Jia}},\ and\
  \bibinfo {author} {\bibfnamefont {M.~Z.}\ \bibnamefont {Hasan}},\ }\bibfield
  {title} {\bibinfo {title} {Discovery of a weyl fermion semimetal and
  topological fermi arcs},\ }\href {https://doi.org/10.1126/science.aaa9297}
  {\bibfield  {journal} {\bibinfo  {journal} {Science}\ }\textbf {\bibinfo
  {volume} {349}},\ \bibinfo {pages} {613} (\bibinfo {year}
  {2015}{\natexlab{a}})},\ \Eprint
  {https://arxiv.org/abs/https://science.sciencemag.org/content/349/6248/613.full.pdf}
  {https://science.sciencemag.org/content/349/6248/613.full.pdf} \BibitemShut
  {NoStop}%
\bibitem [{\citenamefont {Xu}\ \emph {et~al.}(2015{\natexlab{b}})\citenamefont
  {Xu}, \citenamefont {Alidoust}, \citenamefont {Belopolski}, \citenamefont
  {Yuan}, \citenamefont {Bian}, \citenamefont {Chang}, \citenamefont {Zheng},
  \citenamefont {Strocov}, \citenamefont {Sanchez}, \citenamefont {Chang},
  \citenamefont {Zhang}, \citenamefont {Mou}, \citenamefont {Wu}, \citenamefont
  {Huang}, \citenamefont {Lee}, \citenamefont {Huang}, \citenamefont {Wang},
  \citenamefont {Bansil}, \citenamefont {Jeng}, \citenamefont {Neupert},
  \citenamefont {Kaminski}, \citenamefont {Lin}, \citenamefont {Jia},\ and\
  \citenamefont {Zahid~Hasan}}]{Xu_Hasan_2015b}%
  \BibitemOpen
  \bibfield  {author} {\bibinfo {author} {\bibfnamefont {S.-Y.}\ \bibnamefont
  {Xu}}, \bibinfo {author} {\bibfnamefont {N.}~\bibnamefont {Alidoust}},
  \bibinfo {author} {\bibfnamefont {I.}~\bibnamefont {Belopolski}}, \bibinfo
  {author} {\bibfnamefont {Z.}~\bibnamefont {Yuan}}, \bibinfo {author}
  {\bibfnamefont {G.}~\bibnamefont {Bian}}, \bibinfo {author} {\bibfnamefont
  {T.-R.}\ \bibnamefont {Chang}}, \bibinfo {author} {\bibfnamefont
  {H.}~\bibnamefont {Zheng}}, \bibinfo {author} {\bibfnamefont {V.~N.}\
  \bibnamefont {Strocov}}, \bibinfo {author} {\bibfnamefont {D.~S.}\
  \bibnamefont {Sanchez}}, \bibinfo {author} {\bibfnamefont {G.}~\bibnamefont
  {Chang}}, \bibinfo {author} {\bibfnamefont {C.}~\bibnamefont {Zhang}},
  \bibinfo {author} {\bibfnamefont {D.}~\bibnamefont {Mou}}, \bibinfo {author}
  {\bibfnamefont {Y.}~\bibnamefont {Wu}}, \bibinfo {author} {\bibfnamefont
  {L.}~\bibnamefont {Huang}}, \bibinfo {author} {\bibfnamefont {C.-C.}\
  \bibnamefont {Lee}}, \bibinfo {author} {\bibfnamefont {S.-M.}\ \bibnamefont
  {Huang}}, \bibinfo {author} {\bibfnamefont {B.}~\bibnamefont {Wang}},
  \bibinfo {author} {\bibfnamefont {A.}~\bibnamefont {Bansil}}, \bibinfo
  {author} {\bibfnamefont {H.-T.}\ \bibnamefont {Jeng}}, \bibinfo {author}
  {\bibfnamefont {T.}~\bibnamefont {Neupert}}, \bibinfo {author} {\bibfnamefont
  {A.}~\bibnamefont {Kaminski}}, \bibinfo {author} {\bibfnamefont
  {H.}~\bibnamefont {Lin}}, \bibinfo {author} {\bibfnamefont {S.}~\bibnamefont
  {Jia}},\ and\ \bibinfo {author} {\bibfnamefont {M.}~\bibnamefont
  {Zahid~Hasan}},\ }\bibfield  {title} {\bibinfo {title} {Discovery of a weyl
  fermion state with fermi arcs in niobium arsenide},\ }\href
  {https://doi.org/10.1038/nphys3437} {\bibfield  {journal} {\bibinfo
  {journal} {Nature Physics}\ }\textbf {\bibinfo {volume} {11}},\ \bibinfo
  {pages} {748} (\bibinfo {year} {2015}{\natexlab{b}})}\BibitemShut {NoStop}%
\bibitem [{\citenamefont {Lu}\ \emph {et~al.}(2015)\citenamefont {Lu},
  \citenamefont {Wang}, \citenamefont {Ye}, \citenamefont {Ran}, \citenamefont
  {Fu}, \citenamefont {Joannopoulos},\ and\ \citenamefont {Solja{\v
  c}i{\'c}}}]{Lu_Soljacic_2015}%
  \BibitemOpen
  \bibfield  {author} {\bibinfo {author} {\bibfnamefont {L.}~\bibnamefont
  {Lu}}, \bibinfo {author} {\bibfnamefont {Z.}~\bibnamefont {Wang}}, \bibinfo
  {author} {\bibfnamefont {D.}~\bibnamefont {Ye}}, \bibinfo {author}
  {\bibfnamefont {L.}~\bibnamefont {Ran}}, \bibinfo {author} {\bibfnamefont
  {L.}~\bibnamefont {Fu}}, \bibinfo {author} {\bibfnamefont {J.~D.}\
  \bibnamefont {Joannopoulos}},\ and\ \bibinfo {author} {\bibfnamefont
  {M.}~\bibnamefont {Solja{\v c}i{\'c}}},\ }\bibfield  {title} {\bibinfo
  {title} {Experimental observation of weyl points},\ }\href
  {https://doi.org/10.1126/science.aaa9273} {\bibfield  {journal} {\bibinfo
  {journal} {Science}\ }\textbf {\bibinfo {volume} {349}},\ \bibinfo {pages}
  {622} (\bibinfo {year} {2015})},\ \Eprint
  {https://arxiv.org/abs/https://science.sciencemag.org/content/349/6248/622.full.pdf}
  {https://science.sciencemag.org/content/349/6248/622.full.pdf} \BibitemShut
  {NoStop}%
\bibitem [{\citenamefont {Nielsen}\ and\ \citenamefont
  {Ninomiya}(1983)}]{Nielsen_Ninomiya_1983}%
  \BibitemOpen
  \bibfield  {author} {\bibinfo {author} {\bibfnamefont {H.}~\bibnamefont
  {Nielsen}}\ and\ \bibinfo {author} {\bibfnamefont {M.}~\bibnamefont
  {Ninomiya}},\ }\bibfield  {title} {\bibinfo {title} {The adler-bell-jackiw
  anomaly and weyl fermions in a crystal},\ }\href
  {https://doi.org/https://doi.org/10.1016/0370-2693(83)91529-0} {\bibfield
  {journal} {\bibinfo  {journal} {Physics Letters B}\ }\textbf {\bibinfo
  {volume} {130}},\ \bibinfo {pages} {389} (\bibinfo {year}
  {1983})}\BibitemShut {NoStop}%
\bibitem [{\citenamefont {Aji}(2012)}]{Aji_2012}%
  \BibitemOpen
  \bibfield  {author} {\bibinfo {author} {\bibfnamefont {V.}~\bibnamefont
  {Aji}},\ }\bibfield  {title} {\bibinfo {title} {Adler-bell-jackiw anomaly in
  weyl semimetals: Application to pyrochlore iridates},\ }\href
  {https://doi.org/10.1103/PhysRevB.85.241101} {\bibfield  {journal} {\bibinfo
  {journal} {Phys. Rev. B}\ }\textbf {\bibinfo {volume} {85}},\ \bibinfo
  {pages} {241101} (\bibinfo {year} {2012})}\BibitemShut {NoStop}%
\bibitem [{\citenamefont {Zyuzin}\ and\ \citenamefont
  {Burkov}(2012)}]{Zyuzin_Burkov_2012}%
  \BibitemOpen
  \bibfield  {author} {\bibinfo {author} {\bibfnamefont {A.~A.}\ \bibnamefont
  {Zyuzin}}\ and\ \bibinfo {author} {\bibfnamefont {A.~A.}\ \bibnamefont
  {Burkov}},\ }\bibfield  {title} {\bibinfo {title} {Topological response in
  weyl semimetals and the chiral anomaly},\ }\href
  {https://doi.org/10.1103/PhysRevB.86.115133} {\bibfield  {journal} {\bibinfo
  {journal} {Phys. Rev. B}\ }\textbf {\bibinfo {volume} {86}},\ \bibinfo
  {pages} {115133} (\bibinfo {year} {2012})}\BibitemShut {NoStop}%
\bibitem [{\citenamefont {Son}\ and\ \citenamefont
  {Spivak}(2013)}]{Son_Spivak_2013}%
  \BibitemOpen
  \bibfield  {author} {\bibinfo {author} {\bibfnamefont {D.~T.}\ \bibnamefont
  {Son}}\ and\ \bibinfo {author} {\bibfnamefont {B.~Z.}\ \bibnamefont
  {Spivak}},\ }\bibfield  {title} {\bibinfo {title} {Chiral anomaly and
  classical negative magnetoresistance of weyl metals},\ }\href
  {https://doi.org/10.1103/PhysRevB.88.104412} {\bibfield  {journal} {\bibinfo
  {journal} {Phys. Rev. B}\ }\textbf {\bibinfo {volume} {88}},\ \bibinfo
  {pages} {104412} (\bibinfo {year} {2013})}\BibitemShut {NoStop}%
\bibitem [{\citenamefont {Gorbar}\ \emph {et~al.}(2014)\citenamefont {Gorbar},
  \citenamefont {Miransky},\ and\ \citenamefont
  {Shovkovy}}]{Gorbar_Miransky_2014}%
  \BibitemOpen
  \bibfield  {author} {\bibinfo {author} {\bibfnamefont {E.~V.}\ \bibnamefont
  {Gorbar}}, \bibinfo {author} {\bibfnamefont {V.~A.}\ \bibnamefont
  {Miransky}},\ and\ \bibinfo {author} {\bibfnamefont {I.~A.}\ \bibnamefont
  {Shovkovy}},\ }\bibfield  {title} {\bibinfo {title} {Chiral anomaly,
  dimensional reduction, and magnetoresistivity of weyl and dirac semimetals},\
  }\href {https://doi.org/10.1103/PhysRevB.89.085126} {\bibfield  {journal}
  {\bibinfo  {journal} {Phys. Rev. B}\ }\textbf {\bibinfo {volume} {89}},\
  \bibinfo {pages} {085126} (\bibinfo {year} {2014})}\BibitemShut {NoStop}%
\bibitem [{\citenamefont {Burkov}(2015)}]{Burkov_2015}%
  \BibitemOpen
  \bibfield  {author} {\bibinfo {author} {\bibfnamefont {A.~A.}\ \bibnamefont
  {Burkov}},\ }\bibfield  {title} {\bibinfo {title} {Negative longitudinal
  magnetoresistance in dirac and weyl metals},\ }\href
  {https://doi.org/10.1103/PhysRevB.91.245157} {\bibfield  {journal} {\bibinfo
  {journal} {Phys. Rev. B}\ }\textbf {\bibinfo {volume} {91}},\ \bibinfo
  {pages} {245157} (\bibinfo {year} {2015})}\BibitemShut {NoStop}%
\bibitem [{\citenamefont {Lu}\ and\ \citenamefont {Shen}(2017)}]{Lu_Shun_2017}%
  \BibitemOpen
  \bibfield  {author} {\bibinfo {author} {\bibfnamefont {H.-Z.}\ \bibnamefont
  {Lu}}\ and\ \bibinfo {author} {\bibfnamefont {S.-Q.}\ \bibnamefont {Shen}},\
  }\bibfield  {title} {\bibinfo {title} {Quantum transport in topological
  semimetals under magnetic fields},\ }\href
  {https://doi.org/10.1007/s11467-016-0609-y} {\bibfield  {journal} {\bibinfo
  {journal} {Frontiers of Physics}\ }\textbf {\bibinfo {volume} {12}},\
  \bibinfo {pages} {127201} (\bibinfo {year} {2017})}\BibitemShut {NoStop}%
\bibitem [{\citenamefont {Nandy}\ \emph {et~al.}(2017)\citenamefont {Nandy},
  \citenamefont {Sharma}, \citenamefont {Taraphder},\ and\ \citenamefont
  {Tewari}}]{Nandy_Tewari_2017}%
  \BibitemOpen
  \bibfield  {author} {\bibinfo {author} {\bibfnamefont {S.}~\bibnamefont
  {Nandy}}, \bibinfo {author} {\bibfnamefont {G.}~\bibnamefont {Sharma}},
  \bibinfo {author} {\bibfnamefont {A.}~\bibnamefont {Taraphder}},\ and\
  \bibinfo {author} {\bibfnamefont {S.}~\bibnamefont {Tewari}},\ }\bibfield
  {title} {\bibinfo {title} {Chiral anomaly as the origin of the planar hall
  effect in weyl semimetals},\ }\href
  {https://doi.org/10.1103/PhysRevLett.119.176804} {\bibfield  {journal}
  {\bibinfo  {journal} {Phys. Rev. Lett.}\ }\textbf {\bibinfo {volume} {119}},\
  \bibinfo {pages} {176804} (\bibinfo {year} {2017})}\BibitemShut {NoStop}%
\bibitem [{\citenamefont {Das}\ \emph {et~al.}(2020)\citenamefont {Das},
  \citenamefont {Singh},\ and\ \citenamefont {Agarwal}}]{Das_Singh_2020}%
  \BibitemOpen
  \bibfield  {author} {\bibinfo {author} {\bibfnamefont {K.}~\bibnamefont
  {Das}}, \bibinfo {author} {\bibfnamefont {S.~K.}\ \bibnamefont {Singh}},\
  and\ \bibinfo {author} {\bibfnamefont {A.}~\bibnamefont {Agarwal}},\
  }\bibfield  {title} {\bibinfo {title} {Chiral anomalies induced transport in
  weyl metals in quantizing magnetic field},\ }\href
  {https://doi.org/10.1103/PhysRevResearch.2.033511} {\bibfield  {journal}
  {\bibinfo  {journal} {Phys. Rev. Res.}\ }\textbf {\bibinfo {volume} {2}},\
  \bibinfo {pages} {033511} (\bibinfo {year} {2020})}\BibitemShut {NoStop}%
\bibitem [{\citenamefont {Li}\ \emph {et~al.}(2018)\citenamefont {Li},
  \citenamefont {Wang}, \citenamefont {He}, \citenamefont {Wang},\ and\
  \citenamefont {Shen}}]{Li_Shen_2018}%
  \BibitemOpen
  \bibfield  {author} {\bibinfo {author} {\bibfnamefont {H.}~\bibnamefont
  {Li}}, \bibinfo {author} {\bibfnamefont {H.-W.}\ \bibnamefont {Wang}},
  \bibinfo {author} {\bibfnamefont {H.}~\bibnamefont {He}}, \bibinfo {author}
  {\bibfnamefont {J.}~\bibnamefont {Wang}},\ and\ \bibinfo {author}
  {\bibfnamefont {S.-Q.}\ \bibnamefont {Shen}},\ }\bibfield  {title} {\bibinfo
  {title} {Giant anisotropic magnetoresistance and planar hall effect in the
  dirac semimetal ${\mathrm{cd}}_{3}{\mathrm{as}}_{2}$},\ }\href
  {https://doi.org/10.1103/PhysRevB.97.201110} {\bibfield  {journal} {\bibinfo
  {journal} {Phys. Rev. B}\ }\textbf {\bibinfo {volume} {97}},\ \bibinfo
  {pages} {201110} (\bibinfo {year} {2018})}\BibitemShut {NoStop}%
\bibitem [{\citenamefont {Shama}\ \emph {et~al.}(2020)\citenamefont {Shama},
  \citenamefont {Gopal},\ and\ \citenamefont {Singh}}]{Shama_Singh_2020}%
  \BibitemOpen
  \bibfield  {author} {\bibinfo {author} {\bibnamefont {Shama}}, \bibinfo
  {author} {\bibfnamefont {R.}~\bibnamefont {Gopal}},\ and\ \bibinfo {author}
  {\bibfnamefont {Y.}~\bibnamefont {Singh}},\ }\bibfield  {title} {\bibinfo
  {title} {Observation of planar hall effect in the ferromagnetic weyl
  semimetal co3sn2s2},\ }\href
  {https://doi.org/https://doi.org/10.1016/j.jmmm.2020.166547} {\bibfield
  {journal} {\bibinfo  {journal} {Journal of Magnetism and Magnetic Materials}\
  }\textbf {\bibinfo {volume} {502}},\ \bibinfo {pages} {166547} (\bibinfo
  {year} {2020})}\BibitemShut {NoStop}%
\bibitem [{\citenamefont {Li}\ \emph {et~al.}(2023)\citenamefont {Li},
  \citenamefont {Cao}, \citenamefont {Cui}, \citenamefont {Yu},\ and\
  \citenamefont {Yao}}]{Li_Yao_2023}%
  \BibitemOpen
  \bibfield  {author} {\bibinfo {author} {\bibfnamefont {L.}~\bibnamefont
  {Li}}, \bibinfo {author} {\bibfnamefont {J.}~\bibnamefont {Cao}}, \bibinfo
  {author} {\bibfnamefont {C.}~\bibnamefont {Cui}}, \bibinfo {author}
  {\bibfnamefont {Z.-M.}\ \bibnamefont {Yu}},\ and\ \bibinfo {author}
  {\bibfnamefont {Y.}~\bibnamefont {Yao}},\ }\bibfield  {title} {\bibinfo
  {title} {Planar hall effect in topological weyl and nodal-line semimetals},\
  }\href {https://doi.org/10.1103/PhysRevB.108.085120} {\bibfield  {journal}
  {\bibinfo  {journal} {Phys. Rev. B}\ }\textbf {\bibinfo {volume} {108}},\
  \bibinfo {pages} {085120} (\bibinfo {year} {2023})}\BibitemShut {NoStop}%
\bibitem [{\citenamefont {Wei}\ \emph {et~al.}(2023)\citenamefont {Wei},
  \citenamefont {Feng},\ and\ \citenamefont {Weng}}]{Wei_Weng_2023}%
  \BibitemOpen
  \bibfield  {author} {\bibinfo {author} {\bibfnamefont {Y.-W.}\ \bibnamefont
  {Wei}}, \bibinfo {author} {\bibfnamefont {J.}~\bibnamefont {Feng}},\ and\
  \bibinfo {author} {\bibfnamefont {H.}~\bibnamefont {Weng}},\ }\bibfield
  {title} {\bibinfo {title} {Spatial symmetry modulation of planar hall effect
  in weyl semimetals},\ }\href {https://doi.org/10.1103/PhysRevB.107.075131}
  {\bibfield  {journal} {\bibinfo  {journal} {Phys. Rev. B}\ }\textbf {\bibinfo
  {volume} {107}},\ \bibinfo {pages} {075131} (\bibinfo {year}
  {2023})}\BibitemShut {NoStop}%
\bibitem [{\citenamefont {Potter}\ \emph {et~al.}(2014)\citenamefont {Potter},
  \citenamefont {Kimchi},\ and\ \citenamefont
  {Vishwanath}}]{Potter_Vishwanath_2014}%
  \BibitemOpen
  \bibfield  {author} {\bibinfo {author} {\bibfnamefont {A.~C.}\ \bibnamefont
  {Potter}}, \bibinfo {author} {\bibfnamefont {I.}~\bibnamefont {Kimchi}},\
  and\ \bibinfo {author} {\bibfnamefont {A.}~\bibnamefont {Vishwanath}},\
  }\bibfield  {title} {\bibinfo {title} {Quantum oscillations from surface
  fermi arcs in weyl and dirac semimetals},\ }\href
  {https://doi.org/10.1038/ncomms6161} {\bibfield  {journal} {\bibinfo
  {journal} {Nature Communications}\ }\textbf {\bibinfo {volume} {5}},\
  \bibinfo {pages} {5161} (\bibinfo {year} {2014})}\BibitemShut {NoStop}%
\bibitem [{\citenamefont {Zhang}\ \emph {et~al.}(2016)\citenamefont {Zhang},
  \citenamefont {Bulmash}, \citenamefont {Hosur}, \citenamefont {Potter},\ and\
  \citenamefont {Vishwanath}}]{Zhang_Vishwanath_2016}%
  \BibitemOpen
  \bibfield  {author} {\bibinfo {author} {\bibfnamefont {Y.}~\bibnamefont
  {Zhang}}, \bibinfo {author} {\bibfnamefont {D.}~\bibnamefont {Bulmash}},
  \bibinfo {author} {\bibfnamefont {P.}~\bibnamefont {Hosur}}, \bibinfo
  {author} {\bibfnamefont {A.~C.}\ \bibnamefont {Potter}},\ and\ \bibinfo
  {author} {\bibfnamefont {A.}~\bibnamefont {Vishwanath}},\ }\bibfield  {title}
  {\bibinfo {title} {Quantum oscillations from generic surface fermi arcs and
  bulk chiral modes in weyl semimetals},\ }\href
  {https://doi.org/10.1038/srep23741} {\bibfield  {journal} {\bibinfo
  {journal} {Scientific Reports}\ }\textbf {\bibinfo {volume} {6}},\ \bibinfo
  {pages} {23741} (\bibinfo {year} {2016})}\BibitemShut {NoStop}%
\bibitem [{\citenamefont {Moll}\ \emph {et~al.}(2016)\citenamefont {Moll},
  \citenamefont {Nair}, \citenamefont {Helm}, \citenamefont {Potter},
  \citenamefont {Kimchi}, \citenamefont {Vishwanath},\ and\ \citenamefont
  {Analytis}}]{Moll_Analytis_2016}%
  \BibitemOpen
  \bibfield  {author} {\bibinfo {author} {\bibfnamefont {P.~J.~W.}\
  \bibnamefont {Moll}}, \bibinfo {author} {\bibfnamefont {N.~L.}\ \bibnamefont
  {Nair}}, \bibinfo {author} {\bibfnamefont {T.}~\bibnamefont {Helm}}, \bibinfo
  {author} {\bibfnamefont {A.~C.}\ \bibnamefont {Potter}}, \bibinfo {author}
  {\bibfnamefont {I.}~\bibnamefont {Kimchi}}, \bibinfo {author} {\bibfnamefont
  {A.}~\bibnamefont {Vishwanath}},\ and\ \bibinfo {author} {\bibfnamefont
  {J.~G.}\ \bibnamefont {Analytis}},\ }\bibfield  {title} {\bibinfo {title}
  {Transport evidence for fermi-arc-mediated chirality transfer in the dirac
  semimetal cd3as2},\ }\href {https://doi.org/10.1038/nature18276} {\bibfield
  {journal} {\bibinfo  {journal} {Nature}\ }\textbf {\bibinfo {volume} {535}},\
  \bibinfo {pages} {266} (\bibinfo {year} {2016})}\BibitemShut {NoStop}%
\bibitem [{\citenamefont {Ominato}\ and\ \citenamefont
  {Koshino}(2016)}]{Koshino_2016}%
  \BibitemOpen
  \bibfield  {author} {\bibinfo {author} {\bibfnamefont {Y.}~\bibnamefont
  {Ominato}}\ and\ \bibinfo {author} {\bibfnamefont {M.}~\bibnamefont
  {Koshino}},\ }\bibfield  {title} {\bibinfo {title} {Magnetotransport in weyl
  semimetals in the quantum limit: Role of topological surface states},\ }\href
  {https://doi.org/10.1103/PhysRevB.93.245304} {\bibfield  {journal} {\bibinfo
  {journal} {Phys. Rev. B}\ }\textbf {\bibinfo {volume} {93}},\ \bibinfo
  {pages} {245304} (\bibinfo {year} {2016})}\BibitemShut {NoStop}%
\bibitem [{\citenamefont {Kaladzhyan}\ and\ \citenamefont
  {Bardarson}(2019)}]{Kaladzhyan_2019}%
  \BibitemOpen
  \bibfield  {author} {\bibinfo {author} {\bibfnamefont {V.}~\bibnamefont
  {Kaladzhyan}}\ and\ \bibinfo {author} {\bibfnamefont {J.~H.}\ \bibnamefont
  {Bardarson}},\ }\bibfield  {title} {\bibinfo {title} {Quantized fermi arc
  mediated transport in weyl semimetal nanowires},\ }\href
  {https://doi.org/10.1103/PhysRevB.100.085424} {\bibfield  {journal} {\bibinfo
   {journal} {Phys. Rev. B}\ }\textbf {\bibinfo {volume} {100}},\ \bibinfo
  {pages} {085424} (\bibinfo {year} {2019})}\BibitemShut {NoStop}%
\bibitem [{\citenamefont {Breitkreiz}\ and\ \citenamefont
  {Brouwer}(2019)}]{Breitkreiz_2019}%
  \BibitemOpen
  \bibfield  {author} {\bibinfo {author} {\bibfnamefont {M.}~\bibnamefont
  {Breitkreiz}}\ and\ \bibinfo {author} {\bibfnamefont {P.~W.}\ \bibnamefont
  {Brouwer}},\ }\bibfield  {title} {\bibinfo {title} {Large contribution of
  fermi arcs to the conductivity of topological metals},\ }\href
  {https://doi.org/10.1103/PhysRevLett.123.066804} {\bibfield  {journal}
  {\bibinfo  {journal} {Phys. Rev. Lett.}\ }\textbf {\bibinfo {volume} {123}},\
  \bibinfo {pages} {066804} (\bibinfo {year} {2019})}\BibitemShut {NoStop}%
\bibitem [{\citenamefont {Sukhachov}\ \emph {et~al.}(2020)\citenamefont
  {Sukhachov}, \citenamefont {Rakov}, \citenamefont {Teslyk},\ and\
  \citenamefont {Gorbar}}]{Sukhachov_2020}%
  \BibitemOpen
  \bibfield  {author} {\bibinfo {author} {\bibfnamefont {P.~O.}\ \bibnamefont
  {Sukhachov}}, \bibinfo {author} {\bibfnamefont {M.~V.}\ \bibnamefont
  {Rakov}}, \bibinfo {author} {\bibfnamefont {O.~M.}\ \bibnamefont {Teslyk}},\
  and\ \bibinfo {author} {\bibfnamefont {E.~V.}\ \bibnamefont {Gorbar}},\
  }\bibfield  {title} {\bibinfo {title} {Fermi arcs and dc transport in
  nanowires of dirac and weyl semimetals},\ }\href
  {https://doi.org/https://doi.org/10.1002/andp.201900449} {\bibfield
  {journal} {\bibinfo  {journal} {Annalen der Physik}\ }\textbf {\bibinfo
  {volume} {532}},\ \bibinfo {pages} {1900449} (\bibinfo {year} {2020})},\
  \Eprint
  {https://arxiv.org/abs/https://onlinelibrary.wiley.com/doi/pdf/10.1002/andp.201900449}
  {https://onlinelibrary.wiley.com/doi/pdf/10.1002/andp.201900449} \BibitemShut
  {NoStop}%
\bibitem [{\citenamefont {Dwivedi}(2018)}]{Dwivedi_2018}%
  \BibitemOpen
  \bibfield  {author} {\bibinfo {author} {\bibfnamefont {V.}~\bibnamefont
  {Dwivedi}},\ }\bibfield  {title} {\bibinfo {title} {Fermi arc reconstruction
  at junctions between weyl semimetals},\ }\href
  {https://doi.org/10.1103/PhysRevB.97.064201} {\bibfield  {journal} {\bibinfo
  {journal} {Phys. Rev. B}\ }\textbf {\bibinfo {volume} {97}},\ \bibinfo
  {pages} {064201} (\bibinfo {year} {2018})}\BibitemShut {NoStop}%
\bibitem [{\citenamefont {Ishida}\ and\ \citenamefont
  {Liebsch}(2018)}]{Ishida_2018}%
  \BibitemOpen
  \bibfield  {author} {\bibinfo {author} {\bibfnamefont {H.}~\bibnamefont
  {Ishida}}\ and\ \bibinfo {author} {\bibfnamefont {A.}~\bibnamefont
  {Liebsch}},\ }\bibfield  {title} {\bibinfo {title} {Fermi arc engineering at
  the interface between two weyl semimetals},\ }\href
  {https://doi.org/10.1103/PhysRevB.98.195426} {\bibfield  {journal} {\bibinfo
  {journal} {Phys. Rev. B}\ }\textbf {\bibinfo {volume} {98}},\ \bibinfo
  {pages} {195426} (\bibinfo {year} {2018})}\BibitemShut {NoStop}%
\bibitem [{\citenamefont {Murthy}\ \emph {et~al.}(2020)\citenamefont {Murthy},
  \citenamefont {Fertig},\ and\ \citenamefont {Shimshoni}}]{Murthy_2020}%
  \BibitemOpen
  \bibfield  {author} {\bibinfo {author} {\bibfnamefont {G.}~\bibnamefont
  {Murthy}}, \bibinfo {author} {\bibfnamefont {H.~A.}\ \bibnamefont {Fertig}},\
  and\ \bibinfo {author} {\bibfnamefont {E.}~\bibnamefont {Shimshoni}},\
  }\bibfield  {title} {\bibinfo {title} {Surface states and arcless angles in
  twisted weyl semimetals},\ }\href
  {https://doi.org/10.1103/PhysRevResearch.2.013367} {\bibfield  {journal}
  {\bibinfo  {journal} {Phys. Rev. Res.}\ }\textbf {\bibinfo {volume} {2}},\
  \bibinfo {pages} {013367} (\bibinfo {year} {2020})}\BibitemShut {NoStop}%
\bibitem [{\citenamefont {Abdulla}\ \emph {et~al.}(2021)\citenamefont
  {Abdulla}, \citenamefont {Rao},\ and\ \citenamefont
  {Murthy}}]{faruk2021farecon}%
  \BibitemOpen
  \bibfield  {author} {\bibinfo {author} {\bibfnamefont {F.}~\bibnamefont
  {Abdulla}}, \bibinfo {author} {\bibfnamefont {S.}~\bibnamefont {Rao}},\ and\
  \bibinfo {author} {\bibfnamefont {G.}~\bibnamefont {Murthy}},\ }\bibfield
  {title} {\bibinfo {title} {Fermi arc reconstruction at the interface of
  twisted weyl semimetals},\ }\href
  {https://doi.org/10.1103/PhysRevB.103.235308} {\bibfield  {journal} {\bibinfo
   {journal} {Phys. Rev. B}\ }\textbf {\bibinfo {volume} {103}},\ \bibinfo
  {pages} {235308} (\bibinfo {year} {2021})}\BibitemShut {NoStop}%
\bibitem [{\citenamefont {Mathur}\ \emph {et~al.}(2023)\citenamefont {Mathur},
  \citenamefont {Yuan}, \citenamefont {Cheng}, \citenamefont {Kaushik},
  \citenamefont {Robredo}, \citenamefont {Vergniory}, \citenamefont {Cano},
  \citenamefont {Yao}, \citenamefont {Jin},\ and\ \citenamefont
  {Schoop}}]{Mathur2023}%
  \BibitemOpen
  \bibfield  {author} {\bibinfo {author} {\bibfnamefont {N.}~\bibnamefont
  {Mathur}}, \bibinfo {author} {\bibfnamefont {F.}~\bibnamefont {Yuan}},
  \bibinfo {author} {\bibfnamefont {G.}~\bibnamefont {Cheng}}, \bibinfo
  {author} {\bibfnamefont {S.}~\bibnamefont {Kaushik}}, \bibinfo {author}
  {\bibfnamefont {I.}~\bibnamefont {Robredo}}, \bibinfo {author} {\bibfnamefont
  {M.~G.}\ \bibnamefont {Vergniory}}, \bibinfo {author} {\bibfnamefont
  {J.}~\bibnamefont {Cano}}, \bibinfo {author} {\bibfnamefont {N.}~\bibnamefont
  {Yao}}, \bibinfo {author} {\bibfnamefont {S.}~\bibnamefont {Jin}},\ and\
  \bibinfo {author} {\bibfnamefont {L.~M.}\ \bibnamefont {Schoop}},\ }\bibfield
   {title} {\bibinfo {title} {Atomically sharp internal interface in a chiral
  weyl semimetal nanowire},\ }\href
  {https://doi.org/10.1021/acs.nanolett.2c05100} {\bibfield  {journal}
  {\bibinfo  {journal} {Nano Letters}\ }\textbf {\bibinfo {volume} {23}},\
  \bibinfo {pages} {2695} (\bibinfo {year} {2023})}\BibitemShut {NoStop}%
\bibitem [{\citenamefont {Chaou}\ \emph {et~al.}(2023)\citenamefont {Chaou},
  \citenamefont {Dwivedi},\ and\ \citenamefont
  {Breitkreiz}}]{chau2023magnetic}%
  \BibitemOpen
  \bibfield  {author} {\bibinfo {author} {\bibfnamefont {A.~Y.}\ \bibnamefont
  {Chaou}}, \bibinfo {author} {\bibfnamefont {V.}~\bibnamefont {Dwivedi}},\
  and\ \bibinfo {author} {\bibfnamefont {M.}~\bibnamefont {Breitkreiz}},\
  }\bibfield  {title} {\bibinfo {title} {Magnetic breakdown and chiral magnetic
  effect at weyl-semimetal tunnel junctions},\ }\href
  {https://doi.org/10.1103/PhysRevB.107.L241109} {\bibfield  {journal}
  {\bibinfo  {journal} {Phys. Rev. B}\ }\textbf {\bibinfo {volume} {107}},\
  \bibinfo {pages} {L241109} (\bibinfo {year} {2023})}\BibitemShut {NoStop}%
\bibitem [{\citenamefont {Chaou}\ \emph {et~al.}(2024)\citenamefont {Chaou},
  \citenamefont {Dwivedi},\ and\ \citenamefont
  {Breitkreiz}}]{chaou2023quantum}%
  \BibitemOpen
  \bibfield  {author} {\bibinfo {author} {\bibfnamefont {A.~Y.}\ \bibnamefont
  {Chaou}}, \bibinfo {author} {\bibfnamefont {V.}~\bibnamefont {Dwivedi}},\
  and\ \bibinfo {author} {\bibfnamefont {M.}~\bibnamefont {Breitkreiz}},\
  }\bibfield  {title} {\bibinfo {title} {Quantum oscillation signatures of
  interface fermi arcs},\ }\href {https://doi.org/10.1103/PhysRevB.110.035116}
  {\bibfield  {journal} {\bibinfo  {journal} {Phys. Rev. B}\ }\textbf {\bibinfo
  {volume} {110}},\ \bibinfo {pages} {035116} (\bibinfo {year}
  {2024})}\BibitemShut {NoStop}%
\bibitem [{\citenamefont {Buccheri}\ \emph {et~al.}(2022)\citenamefont
  {Buccheri}, \citenamefont {Egger},\ and\ \citenamefont
  {De~Martino}}]{buccheri2022transport}%
  \BibitemOpen
  \bibfield  {author} {\bibinfo {author} {\bibfnamefont {F.}~\bibnamefont
  {Buccheri}}, \bibinfo {author} {\bibfnamefont {R.}~\bibnamefont {Egger}},\
  and\ \bibinfo {author} {\bibfnamefont {A.}~\bibnamefont {De~Martino}},\
  }\bibfield  {title} {\bibinfo {title} {Transport, refraction, and interface
  arcs in junctions of weyl semimetals},\ }\href
  {https://doi.org/10.1103/PhysRevB.106.045413} {\bibfield  {journal} {\bibinfo
   {journal} {Phys. Rev. B}\ }\textbf {\bibinfo {volume} {106}},\ \bibinfo
  {pages} {045413} (\bibinfo {year} {2022})}\BibitemShut {NoStop}%
\bibitem [{\citenamefont {Tchoumakov}\ \emph {et~al.}(2021)\citenamefont
  {Tchoumakov}, \citenamefont {Bujnowski}, \citenamefont {Noky}, \citenamefont
  {Gooth}, \citenamefont {Grushin},\ and\ \citenamefont
  {Cayssol}}]{tchoumakov201conservation}%
  \BibitemOpen
  \bibfield  {author} {\bibinfo {author} {\bibfnamefont {S.}~\bibnamefont
  {Tchoumakov}}, \bibinfo {author} {\bibfnamefont {B.}~\bibnamefont
  {Bujnowski}}, \bibinfo {author} {\bibfnamefont {J.}~\bibnamefont {Noky}},
  \bibinfo {author} {\bibfnamefont {J.}~\bibnamefont {Gooth}}, \bibinfo
  {author} {\bibfnamefont {A.~G.}\ \bibnamefont {Grushin}},\ and\ \bibinfo
  {author} {\bibfnamefont {J.}~\bibnamefont {Cayssol}},\ }\bibfield  {title}
  {\bibinfo {title} {Conservation of chirality at a junction between two weyl
  semimetals},\ }\href {https://doi.org/10.1103/PhysRevB.104.125308} {\bibfield
   {journal} {\bibinfo  {journal} {Phys. Rev. B}\ }\textbf {\bibinfo {volume}
  {104}},\ \bibinfo {pages} {125308} (\bibinfo {year} {2021})}\BibitemShut
  {NoStop}%
\bibitem [{\citenamefont {de~Sousa}\ \emph {et~al.}(2021)\citenamefont
  {de~Sousa}, \citenamefont {Ascencio}, \citenamefont {Haney}, \citenamefont
  {Wang},\ and\ \citenamefont {Low}}]{sousa2021gigantic}%
  \BibitemOpen
  \bibfield  {author} {\bibinfo {author} {\bibfnamefont {D.~J.~P.}\
  \bibnamefont {de~Sousa}}, \bibinfo {author} {\bibfnamefont {C.~O.}\
  \bibnamefont {Ascencio}}, \bibinfo {author} {\bibfnamefont {P.~M.}\
  \bibnamefont {Haney}}, \bibinfo {author} {\bibfnamefont {J.~P.}\ \bibnamefont
  {Wang}},\ and\ \bibinfo {author} {\bibfnamefont {T.}~\bibnamefont {Low}},\
  }\bibfield  {title} {\bibinfo {title} {Gigantic tunneling magnetoresistance
  in magnetic weyl semimetal tunnel junctions},\ }\href
  {https://doi.org/10.1103/PhysRevB.104.L041401} {\bibfield  {journal}
  {\bibinfo  {journal} {Phys. Rev. B}\ }\textbf {\bibinfo {volume} {104}},\
  \bibinfo {pages} {L041401} (\bibinfo {year} {2021})}\BibitemShut {NoStop}%
\bibitem [{\citenamefont {Groth}\ \emph {et~al.}(2014)\citenamefont {Groth},
  \citenamefont {Wimmer}, \citenamefont {Akhmerov},\ and\ \citenamefont
  {Waintal}}]{groth2023kwant}%
  \BibitemOpen
  \bibfield  {author} {\bibinfo {author} {\bibfnamefont {C.~W.}\ \bibnamefont
  {Groth}}, \bibinfo {author} {\bibfnamefont {M.}~\bibnamefont {Wimmer}},
  \bibinfo {author} {\bibfnamefont {A.~R.}\ \bibnamefont {Akhmerov}},\ and\
  \bibinfo {author} {\bibfnamefont {X.}~\bibnamefont {Waintal}},\ }\bibfield
  {title} {\bibinfo {title} {Kwant: a software package for quantum transport},\
  }\href {https://doi.org/10.1088/1367-2630/16/6/063065} {\bibfield  {journal}
  {\bibinfo  {journal} {New J. Phys.}\ }\textbf {\bibinfo {volume} {16}},\
  \bibinfo {pages} {063065} (\bibinfo {year} {2014})}\BibitemShut {NoStop}%
\bibitem [{\citenamefont {Kobayashi}\ \emph {et~al.}(2018)\citenamefont
  {Kobayashi}, \citenamefont {Ominato},\ and\ \citenamefont
  {Nomura}}]{Kobayashi_2018}%
  \BibitemOpen
  \bibfield  {author} {\bibinfo {author} {\bibfnamefont {K.}~\bibnamefont
  {Kobayashi}}, \bibinfo {author} {\bibfnamefont {Y.}~\bibnamefont {Ominato}},\
  and\ \bibinfo {author} {\bibfnamefont {K.}~\bibnamefont {Nomura}},\
  }\bibfield  {title} {\bibinfo {title} {Helicity-protected domain-wall
  magnetoresistance in ferromagnetic weyl semimetal},\ }\href
  {https://doi.org/10.7566/JPSJ.87.073707} {\bibfield  {journal} {\bibinfo
  {journal} {Journal of the Physical Society of Japan}\ }\textbf {\bibinfo
  {volume} {87}},\ \bibinfo {pages} {073707} (\bibinfo {year} {2018})},\
  \Eprint {https://arxiv.org/abs/https://doi.org/10.7566/JPSJ.87.073707}
  {https://doi.org/10.7566/JPSJ.87.073707} \BibitemShut {NoStop}%
\bibitem [{\citenamefont {Kim}\ \emph {et~al.}(2017)\citenamefont {Kim},
  \citenamefont {Ryoo},\ and\ \citenamefont {Park}}]{Kim_breakdown_2017}%
  \BibitemOpen
  \bibfield  {author} {\bibinfo {author} {\bibfnamefont {P.}~\bibnamefont
  {Kim}}, \bibinfo {author} {\bibfnamefont {J.~H.}\ \bibnamefont {Ryoo}},\ and\
  \bibinfo {author} {\bibfnamefont {C.-H.}\ \bibnamefont {Park}},\ }\bibfield
  {title} {\bibinfo {title} {Breakdown of the chiral anomaly in weyl semimetals
  in a strong magnetic field},\ }\href
  {https://doi.org/10.1103/PhysRevLett.119.266401} {\bibfield  {journal}
  {\bibinfo  {journal} {Phys. Rev. Lett.}\ }\textbf {\bibinfo {volume} {119}},\
  \bibinfo {pages} {266401} (\bibinfo {year} {2017})}\BibitemShut {NoStop}%
\bibitem [{\citenamefont {Chan}\ and\ \citenamefont
  {Lee}(2017)}]{Chan_emergence_2017}%
  \BibitemOpen
  \bibfield  {author} {\bibinfo {author} {\bibfnamefont {C.-K.}\ \bibnamefont
  {Chan}}\ and\ \bibinfo {author} {\bibfnamefont {P.~A.}\ \bibnamefont {Lee}},\
  }\bibfield  {title} {\bibinfo {title} {Emergence of gapped bulk and metallic
  side walls in the zeroth landau level in dirac and weyl semimetals},\ }\href
  {https://doi.org/10.1103/PhysRevB.96.195143} {\bibfield  {journal} {\bibinfo
  {journal} {Phys. Rev. B}\ }\textbf {\bibinfo {volume} {96}},\ \bibinfo
  {pages} {195143} (\bibinfo {year} {2017})}\BibitemShut {NoStop}%
\bibitem [{\citenamefont {Abdulla}\ \emph {et~al.}(2022)\citenamefont
  {Abdulla}, \citenamefont {Das}, \citenamefont {Rao},\ and\ \citenamefont
  {Murthy}}]{Abdulla_timerev_2022}%
  \BibitemOpen
  \bibfield  {author} {\bibinfo {author} {\bibfnamefont {F.}~\bibnamefont
  {Abdulla}}, \bibinfo {author} {\bibfnamefont {A.}~\bibnamefont {Das}},
  \bibinfo {author} {\bibfnamefont {S.}~\bibnamefont {Rao}},\ and\ \bibinfo
  {author} {\bibfnamefont {G.}~\bibnamefont {Murthy}},\ }\bibfield  {title}
  {\bibinfo {title} {{Time-reversal-broken Weyl semimetal in the Hofstadter
  regime}},\ }\href {https://doi.org/10.21468/SciPostPhysCore.5.1.014}
  {\bibfield  {journal} {\bibinfo  {journal} {SciPost Phys. Core}\ }\textbf
  {\bibinfo {volume} {5}},\ \bibinfo {pages} {014} (\bibinfo {year}
  {2022})}\BibitemShut {NoStop}%
\bibitem [{\citenamefont {Abdulla}(2024)}]{Abdulla_pairwise_2024}%
  \BibitemOpen
  \bibfield  {author} {\bibinfo {author} {\bibfnamefont {F.}~\bibnamefont
  {Abdulla}},\ }\bibfield  {title} {\bibinfo {title} {Pairwise annihilation of
  weyl nodes induced by magnetic fields in the hofstadter regime},\ }\href
  {https://doi.org/10.1103/PhysRevB.109.155142} {\bibfield  {journal} {\bibinfo
   {journal} {Phys. Rev. B}\ }\textbf {\bibinfo {volume} {109}},\ \bibinfo
  {pages} {155142} (\bibinfo {year} {2024})}\BibitemShut {NoStop}%
\end{thebibliography}%

\end{document}